\documentclass[10pt,journal,compsoc]{IEEEtran}
\pdfoutput=1

\usepackage{graphicx}
\usepackage{float}
\usepackage{url}
\usepackage{mathrsfs}
\usepackage{amsthm}

\usepackage{amsfonts}
\usepackage{color}
\usepackage{multirow}
\usepackage{caption}
\usepackage{graphicx} 
\usepackage{epstopdf}
\usepackage{algorithm}
\usepackage{algorithmic}
\usepackage{amsmath}
\usepackage{subfigure}
\usepackage[square, comma, sort&compress, numbers]{natbib}
\usepackage{amssymb}
\usepackage[english]{babel}
\usepackage{blindtext}
\usepackage{graphics}
\usepackage{makecell}
\usepackage{epsfig}
\usepackage{bbding}
\usepackage{booktabs}

\newtheorem{theorem}{Theorem}[section]
\newtheorem{lemma}[theorem]{Lemma}

\usepackage[utf8]{inputenc}

\ifCLASSINFOpdf

\else

\fi

\begin{document}

\title{Bodyless Block Propagation: TPS Fully Scalable Blockchain with Pre-Validation}

\author{\IEEEauthorblockN{ Chonghe Zhao, Shengli Zhang, \IEEEmembership{Senior Member, IEEE}, Taotao Wang, \IEEEmembership{Member, IEEE}, Soung Chang Liew, \IEEEmembership{Fellow, IEEE}}
	\IEEEcompsocitemizethanks{\IEEEcompsocthanksitem Chonghe Zhao is with the College of Electronics and Information Engineering, Shenzhen University, Shenzhen 518060, China. \protect\\ E-mail: zhaochonghe\_szu@163.com
		
		\IEEEcompsocthanksitem Shengli Zhang and Taotao Wang are with the College of Electronics and Information Engineering, Shenzhen University, Shenzhen 518060, China, and also with MAXE LAB. E-mail: zsl@szu.edu.cn; ttwang@szu.edu.cn
		
		\IEEEcompsocthanksitem Soung Chang Liew is with the Department of Information Engineering, The Chinese University of Hong Kong, Hong Kong SAR, China, and also with MAXE LAB. E-mail: soung@ie.cuhk.edu.hk
	}

\thanks{Corresponding author: Shengli Zhang}
}








\IEEEtitleabstractindextext{
\begin{abstract}
 
Despite numerous prior attempts to boost transaction per second (TPS) of blockchain systems, many sacrifice decentralization and security. This paper proposes a bodyless block propagation (BBP) scheme for which the blockbody is not validated and transmitted during block propagation, to increase TPS without compromising security. Nodes in the blockchain network anticipate the transactions and their ordering in the next upcoming block so that these transactions can be pre-executed and pre-validated before the block is born. For a network with $N$ nodes, our theoretical analysis reveals that BBP can improve TPS scalability from $O(1/log(N))$ to $O(1)$. 

Ensuring consensus on the next block's transaction content is crucial. We propose a transaction selection, ordering, and synchronization algorithm to drive this consensus. To address the undetermined Coinbase address issue, we further present an algorithm for such unresolvable transactions, ensuring a consistent and TPS-efficient scheme. With BBP, most transactions require neither validation nor transmission during block propagation, liberating system from transaction-block dependencies and rendering TPS scalable. Both theoretical analysis and experiments underscore BBP's potential for full TPS scalability. Experimental results reveal a 4x reduction in block propagation time compared to Ethereum blockchain, with TPS performance being limited by node hardware rather than block propagation.

\end{abstract}

}

\maketitle

\IEEEpeerreviewmaketitle

\section{Introduction}\label{introduction}


In 2008, Satoshi Nakamoto proposed Bitcoin, a peer-to-peer (P2P) electronic cash system \cite{bitcoin}. Bitcoin fundamentally challenged the role of traditional banking systems by enabling decentralized money transfers over a network consisting of untrusted nodes. The underlying enabling technology of Bitcoin came to be known as blockchain. Blockchain is a secure, verifiable, and tamper-proof distributed ledger technology for recording transactions. The design of Bitcoin's blockchain integrates the advances of cryptography, distributed systems, and P2P networks. Following this genesis design, many efforts have been dedicated to the investigations and deployments of other blockchain systems. Among them, perhaps the most successful is Ethereum \cite{ethereum,wood2014ethereum}. Ethereum introduces smart contracts to enable the execution of Turing-complete computing tasks, greatly extending the application and usefulness of blockchain. For example, the currently in-vogue Decentralized Finance (DeFi) \cite{mev_1}, Non-Fungible Token (NFT) \cite{NFT_2}, and Metaverse \cite{metaverse} are all built on blockchain, which is becoming a service platform. 

A weakness of the present blockchains is their low data processing capability, as measured by transactions per second (TPS). For example, as shown in \cite{ethereum_tps,bitcoinTPS}, the TPS of Bitcoin and Ethereum are 7 and 15, respectively, significantly lower than that of centralized systems. The low TPS cannot meet the needs of large-scale applications, imposing a fundamental limit on the blockchain's utility. 
Moreover, with the deployment of DeFi and NFT applications, the Ethereum network is increasingly congested. According to the Ethereum block explorer, Etherscan \cite{etherscan}, the average number of pending transactions on the network has accumulated to 138,000. Improving blockchain's TPS is of paramount importance.

\begin{figure}[tp]
	\centering
        \includegraphics[width=\linewidth]{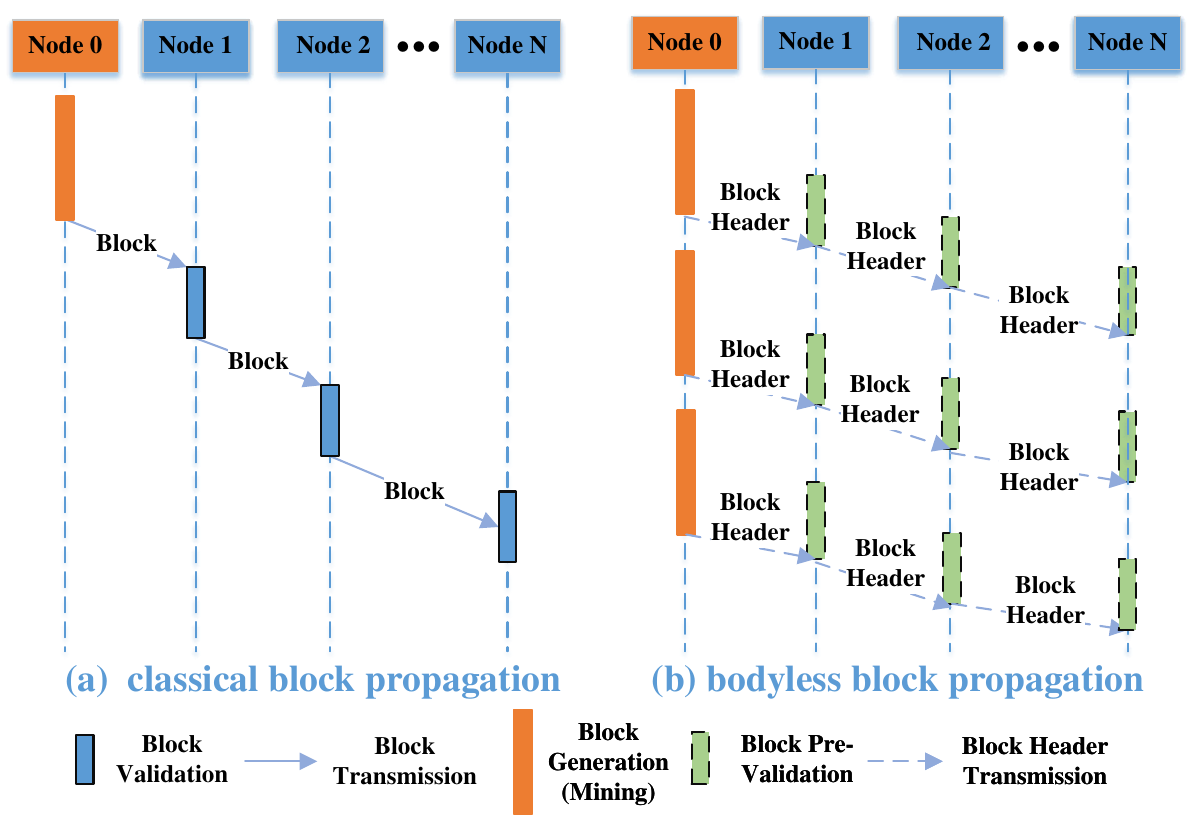}
	\caption{Illustration of block propagations: (a) classical block propagation; (b) bodyless block propagation.}
	\label{BBPidea}
\end{figure}

An intuitive but naïve way to improve TPS is to enlarge the block size so that each block can carry more transactions. But large blocks decelerate their propagation on the network, thereby compromising the blockchain's security and integrity \cite{blocksecurity}.  
Fig. \ref{BBPidea}(a) illustrates the time spent in block propagation. In Fig. \ref{BBPidea}(a), Node 0 is the miner that has mined a block and propagated the block to all other nodes over several communication hops. To fend off invalid blocks, each node must verify the received block before forwarding it to neighbor nodes. Thus, block propagation time consists primarily of block validation time and block transmission time on the network.

Accordingly, we can accelerate block propagation by decreasing i) block transmission time and ii) block validation time. To lower block transmission time, Bitcoin Improvement Proposal (BIP) 152 proposed to propagate compact blocks in place of full blocks. In compact blocks, transaction hashes replace the transactions as in full blocks \cite{compact,hcb}, thereby compressing the block size. If a node receives a compact block and cannot locate some transactions corresponding to the transaction hashes from its local transaction pool, the node needs to request the missing transactions from other nodes. In situations where the transactions at different nodes' transaction pools differentiate vastly, there is a high probability of the time-consuming request-response process and retarding block propagation. Notably, while compact block works well in Bitcoin, it performs poorly in Ethereum because the shorter inter-block interval in Ethereum leads to substantial transaction differences among different nodes' transaction pools. The transactions in a new block may not have time to propagate to all nodes yet by the time the new block is born\footnote{Section \ref{measurement} of this paper provides experimental evidence to the failure of compact block in Ethereum.}. 

Assuming a similar idea as compact blocks, \cite{xthinblock, graphene} improved the block propagation efficiency using a Bloom Filter and Invertible Bloom Lookup Tables (IBLT) for quick detection of the missing transactions in the compressed blocks. However, the block size still increases with the number of transaction hashes in these schemes, and nodes still need extra time to synchronize missing transactions in the block. In conclusion, compact blocks may alleviate the block propagation problem under certain conditions, but they fall short in other situations and their TPS scalability is limited. 

Besides, another way to speed up block propagation is to reduce block validation time. As pointed out in \cite{theoreticalBitcoin}, block validation time cannot be neglected and is the bottleneck of block propagation for large block sizes. To shorten block validation time, \cite{reducingfork} proposed to probabilistically validate received new blocks. Not validating all blocks, however, may compromise securities. Also, selective block validation still does not solve the aforementioned missing transaction issue. 

This work puts forth a novel bodyless block propagation (BBP) scheme for Ethereum-like blockchains. BBP nails block propagation time to a constant level that does not increase with the number of transactions in a block, thereby allowing many transactions in a block. 
The key is that BBP only propagates the blockheader, ridding the dependence of block propagation speed on the number of transactions in the blockbody. A block can then pack any number of transactions in the unpropagated blockbody to boost TPS. As shown in our analysis, any blockchain system, with a blockbody proportional to the number of transactions, can only achieve scalability of $ O(1/log(N))$ ($N$ is the number of nodes in the network), while our BBP can achieve the scalability upper bound of $O (1)$.

As shown in Fig. \ref{BBPidea}(b), the underpinning of BBP is that each node anticipates the transactions in the upcoming next block and pre-packs a blockbody based on the anticipated transactions. The node pre-executes and pre-validates the anticipated transactions before the next block is mined. To the extent that the anticipated transactions of all nodes are the same and that the miner of the next block also embeds the same transactions in the block, then the block validation has already been done by all nodes by the time the next block is mined. Only the blockheader needs to be propagated. As soon as it receives the block header, a node can verify that the validation information in the header is the same as the validation information of the pre-packed blockbody. 

It is more challenging to design and implement the basic BBP scheme in a real blockchain system, e.g., Ethereum. This paper puts forth solutions and verifies their effectiveness. The main contributions are as follows:
\begin{itemize}[left=0pt]
\item  We put forth the basic idea of BBP, where each node pre-packs and pre-validates the next blockbody so that the miner only needs to broadcast the blockheader. With a theoretical analysis, we prove that BBP can achieve a scalability of $O(1)$, while all the current blockchain schemes can only achieve $O(1/log(N))$. 
\item To implement BBP in Ethereum, we put forth a Time-specific transaction Selection and Ordering (TSO) algorithm to disambiguate and align the transaction selection and ordering at all nodes. Although nodes can independently and freely select transactions for blocks, rational nodes would not do so because they can maximize their GAS revenues and save computing costs by executing TSO. In addition, we propose a pre-packed blockbody synchronization protocol with which the nodes exchange pre-packed blockbody information with their neighbors to synchronize their blockbody securely.
\item To improve BBP in Ethereum, we put forth a pre-validation algorithm to accelerate block propagation. For block validation in Ethereum and most blockchains that support smart contracts, nodes need to execute the transactions to compute a updated global state. 
Since the execution order of the transactions matters to the global state, by convention the transactions in a block are executed sequentially. However, the Coinbase transaction information of the block is not available yet before the arrival of the new block. To achieve effective pre-validation, we propose an algorithm to identify the transaction chain that involves the Coinbase transaction. Then, the pre-validation algorithm can pre-validate the block to the largest possible extent, leaving only transactions that depend on Coinbase to be validated during block propagation.
\item We implemented BBP and conducted experiments over a large-scale blockchain network with many nodes to evaluate its performance. We compare BBP with other block propagation schemes based on the experimental results. The experiment results show that BBP has the least block propagation time. Compared with the current protocol of Ethereum, BBP reduces the block propagation time by 4x. Importantly, our work shows that BBP has a constant-level block propagation time for blocks with modest to large numbers of transactions, thereby demonstrating the full TPS scalability of BBP. 

\end{itemize}

The rest of this paper is organized as follows. Section \ref{theoretical} analyzes the theoretical scalability of our basic BBP scheme. Section \ref{measurement_main} discusses the design of a practical BBP scheme based on Ethereum. Section \ref{BBPDesign} follows by elaborating on the systematic design of BBP. Section \ref{expeva} analyzes the block propagation delays of different protocols and discusses our experimental results. Section \ref{RELATEWORK} overviews related work. Section \ref{CONCLUSION} concludes this work.

\textbf{Ethics Statement:} This work does not raise any ethical issues.

\section{Basic BBP Scheme and Scalability Analysis}\label{theoretical}
\subsection{Blockchain TPS Model}
Consider a blockchain system consisting of $N$ peers. In a typical blockchain, there are mainly two procedures. The first procedure is the transaction distribution procedure where each peer randomly generates transactions with the same rate and broadcasts the transactions over the peer-to-peer network with a gossip protocol. Denote the overall transaction generation rate of the system by $\lambda_1(N)$. The first procedure is an all-to-all broadcasting over the P2P network. Denote the expected time for broadcasting one transaction to most of the peers by $d_1(N)$. The second procedure is the consensus procedure where the peers will reach an agreement to each new block. Therefore, a new block needs to be broadcast and recognized by most of the $N$ peers. Denote the overall block generation rate by $\lambda_2(N)$. The second procedure includes at least a one-to-all broadcast (in the case of forks a few nodes generate and broadcast the new block at the same time), and denotes the expected time for one block to be recognized by most of the nodes by $d_2(N)$. Suppose there are $k$ transactions in one block on average in a stable blockchain system. 
The TPS is $ TPS = min(\lambda_1(N),k*\lambda_2(N))$. 

\subsection{Scalability Analysis for Typical Blockchains}\label{Scalability_typical_blockchain}
In a P2P network with $N$ peers adopting a gossip protocol to broadcast messages, Ref.\cite{multipleTxs} and Ref.\cite{gossip_ds} demonstrate broadcasting a message to all peers can be finished in $O(log(N))$ time unit. 
Suppose each link bandwidth is one transaction per time unit in a typical blockchain, and the expected time for broadcasting one transaction to $1-\epsilon$ (where $\epsilon$ is a small value) portion of peers can be rewritten in following lemma.  
\begin{lemma} \label{lemma1}
    Broadcasting one transaction to $1-\epsilon$ portion of the nodes can be finished in $O(log(N)+log(1/\epsilon))$ time units.
\end{lemma}

For the block broadcasting in the consensus procedure, ignoring the constant overhead in a block, each link can transmit one block within $k/c$ time unit, where the constant factor $c>1$ comes from some compression techniques. According to Lemma \ref{lemma1}, the time for broadcasting one block to most of the nodes is $d_2(N)=O(k/c*log(N))$. Then, the block generation rate should be no more than $\lambda_2(N) = O(c/(k*log(N)))$ for stability. At best, $O(c/log(N))$ transactions can be broadcast along block per time unit, and the blockchain will keep stable with typical consensus algorithms \cite{stability}.

Ref.\cite{multipleTxs} and Ref.\cite{gossip} also demonstrate broadcasting $N$ messages to $N$ peers in $O (N+log(N))$ time units, where each peer has a unique message at the beginning. It implies that each link transmission is efficient (useful) for all-to-all broadcasts. Thus, with a sufficient transaction generation rate at blockchain nodes, the expected time for broadcasting $N$ transactions over a P2P network, as the transaction broadcasting procedure, can be given by the following lemma. 
\begin{lemma}
Broadcasting $N$ transactions over the blockchain nodes can be finished in $O(N+log(N))$ time units.
\end{lemma}

As a result, the maximum transaction generation rate, only considering the transaction broadcasting procedure, should be no more than $\lambda_1(N)=O(N/(N+log(N)))$ for stability.

\begin{theorem}
The largest supportable TPS of the classical blockchain system is $k*\lambda_2(N)=O(c/log(N))$, which means that the classical blockchain is not scalable and its bottleneck is the block propagation time during the consensus procedure. 
\end{theorem}
 
\subsection{Basic BBP Scheme}
According to the analysis in Section \ref{Scalability_typical_blockchain}, the only way to increase the TPS in the consensus procedure is to insert more than $log(N)$ transactions into one block without increasing broadcasting time (i.e., maintaining block broadcasting time within $O(log(N))$ time units).  So we propose bodyless block propagation scheme (BBP), and its basic idea is as follows.

In BBP, when a block is mined, only its blockheader is broadcast, and the blockbody is not broadcast at all. Each peer will pre-assemble the blockbody of the new block. 
For all peers to assemble the same blockbody, all pending transactions, with a timestamp earlier than $d_1(N)$, are packed according to the order of their timestamps.  When a peer receives a BBP blockheader,  its pre-assembled blockbody matches the BBP blockheader with high probability and it can reconstruct the complete block. After verification, the BBP blockheader is continuously forwarded. 

Therefore, BBP only broadcasts the blockheader in the consensus procedure. Its broadcasting time is constant with any number of transactions inside one block.

\subsection{Scalability Analysis for BBP Scheme}

\begin{lemma}
   In the all-to-all transaction broadcasting procedure, each transaction can be distributed to at least $1-\epsilon$ of the peers within $d_1(N)=O(log(N)+log(1/\epsilon)$, where $\epsilon<<1$ is a small value.
\end{lemma}

\begin{proof}
    According to Lemma 2.1, the delay is $d_1(N)=O(log(N)+log(1/\epsilon))$ to distribute one message/transaction over $1-\epsilon$ of the peers. For the continuous all-to-all transaction broadcasting procedure, the delay may be extended to waiting for other transactions in queue to broadcast. With a stable transaction generation rate $\lambda_1$, the average queue length at each peer is bounded by a constant, $c_1$.  Then $d_1(N)=O(c_1*log(N)+c_1*log(1/\epsilon))=O(log(N)+log(1/\epsilon))$.    
\end{proof}
If we set $\epsilon=1/log^2(N)$, then the needed transaction broadcasting time is $O(log(N)+2log(log(N))$.

\begin{lemma}
  For the BBP block broadcasting, each block can be distributed to $1-\delta$ of the peers within $d_2(N)=O(log(N))$, where the small value $\delta \leq O(1/log(N))$.
\end{lemma}  

\begin{proof}
    The BBP blockheader 
    distribution is the same as the traditional scheme, ignoring the peers that do not have the same blockbody. Then, the distribution time is $O(log(N))$. We now show that the number of peers assembling a different blockbody, i.e., the peers do not have all transactions in the blockbody, is less than $\delta$. 
    
   With a stable transaction generation rate $\lambda_1$, the average transactions in one block are $k=O(N*log(N)/(N+log(N)))=O(log(N))$.

    Since every transaction is distributed to at least $1-\epsilon$ of the peers, each peer has a probability of $\Lambda=1-(1-\epsilon)^k$ that it cannot construct the same blockbody.  According to Taylor Series, $\Lambda\leq1-(1-k\epsilon)$. By setting $\epsilon=1/log^2(N)$, we have
    
    \begin{equation}
	\setlength{\abovedisplayskip}{3pt}
	\setlength{\belowdisplayskip}{3pt}
 	\begin{split}
            \Lambda\leq O(1-(1-1/log^2(N))^{log(N)}) \\
            =O(1-(1-log(N)/log^2(N))) 
            = O(1/log(N))
	\end{split}
\end{equation}
   
    Thus, the block can be broadcasted to at least $1-\delta$ of the peers within $O(log(N))$ time units.   
\end{proof}

By combining Lemma 2.4 and Lemma 2.5, we obtain the scalability theorem for BBP as follows.
\begin{theorem}
    The largest supportable TPS of the BBP blockchain system is  $\lambda_1(N)=O (N/(N+log(N)))=O (1)$, which means that the BBP blockchain system is scalable and its bottleneck is the transaction broadcast procedure. 
\end{theorem}
\section{Apply BBP in Ethereum}\label{measurement_main}
While the basic BBP scheme can achieve scalability in theory, its applicability in real blockchain systems, e.g., Ethereum, needs further discussion. This section delves into experiments on block validation and transmission time in Ethereum.  The results highlight both the prospects and technical challenges in designing a practical BBP scheme for Ethereum. 


\subsection{Measurement and Observation on Ethereum}\label{measurement}
\subsubsection{Block propagation time on Ethereum TPS}
We can write the TPS in the Ethereum blockchain as
\begin{equation}
	\setlength{\abovedisplayskip}{3pt}
	\setlength{\belowdisplayskip}{3pt}
	\label{TPS1}
	\begin{split}
		TPS{\rm{ = }}\frac{{{s_{{b}}}}}{{{s_t}{t_g}}} = \frac{{{n_t}}}{{{t_g}}}
	\end{split}
\end{equation}
where $s_b$ is the average size of a full block, $s_t$  is the average transaction size,  $t_g$ is the average time interval between two successive blocks, and $n_t = s_b / s_t $  is the number of transactions in the block. In the current Ethereum, a block contains 200 transactions on average and $t_g$ is around 13-15s, and thus the TPS of Ethereum is less than 15. 

From eq. (\ref{TPS1}), increasing $s_b$ or decreasing  $t_g$ is a simple way to improve TPS. However, merely increasing $s_b$  or decreasing $t_g$ does not work. 
Ensuring security mandates that blocks be propagated to nearly all nodes in the blockchain network, typically 90\%. 
We denote the time to propagate a block from the miner node to the 90\% of the nodes in the blockchain network by $t_l$. The probability of the occurrence of stale blocks (uncle blocks in Ethereum) is denoted by $p_f$. The relationship between  $t_l$  and  $p_f$  can be approximated as \cite{bloxroute}:
\begin{equation}
	\setlength{\abovedisplayskip}{3pt}
	\setlength{\belowdisplayskip}{3pt}
	\label{fork}
	\begin{split}
		{p_f} = 1 - {e^{-\frac{{ {t_l}}}{{{t_g}}}}}
	\end{split}
\end{equation}
In current blockchain block propagation protocols, $t_l$ and $n_t$, exhibit a linear relationship,  ${t_l} \approx k{n_t}$, where $k$ is a constant. Using this relationship, we can rewrite eq. (\ref{fork}) as:
\begin{equation}
	\setlength{\abovedisplayskip}{3pt}
	\setlength{\belowdisplayskip}{3pt}
	\label{TPS}
	\begin{split}
		{p_f} = 1 - {e^{-\frac{{ {t_l}}}{{{t_g}}}}} = 1 - {e^{-\frac{{k{n_t}}}{{{t_g}}}}} = 1 - {e^{ - k*TPS}}
	\end{split}
\end{equation}

From eq. (\ref{TPS}), we can see that, with a fixed $t_g$, the increase of $n_t$ would increase  $t_l$ and $p_f$, weakening the security of the blockchain, since the blockchain system is more vulnerable to various attacks (e.g., double-spend attacks \cite{37aa}) when $p_f$ is large. In other words, there is always a tradeoff between TPS and security with the current blockchain network, in which the block propagation time increases with the block size\footnote{Although this result is presented under the context of PoW \cite{pow_1}, it also applies to other consensus, including PoS, PoA, PBFT, so long as block propagation time increases with the block size in the consensus process.}. We are thus led to the following design principle to tackle the TPS scalability problem without sacrificing security: 

\noindent\textbf{\emph{Design Principle:}} When and only when the block propagation time is independent of the number of transactions in a block, the TPS of blockchain can be scaled without compromising system security.


\subsubsection{Measurement on block validation time}
In Ethereum, a node must validate a newly received block before fully forwarding it to neighbors to avoid propagating illegal blocks. As such, the block validation time directly impacts the efficiency of the block consensus process.

The specific block validation process in Ethereum includes three steps. Firstly, an Ethereum node will validate the correctness of the blockheader, including the timestamp and the nonce value. If the blockheader is correct, the Ethereum node will secondly check the correctness of the new global state root value by sequentially executing all transactions in the blockbody. Finally, the Ethereum node stores the new global state and the new block in its local database if the second step is passed. The block will be discarded if any step fails in the process of block validation.

As discussed above, the block validation time consists of three components: header validation time,  transaction execution time, and data storage time. To measure the three time components in Ethereum, we set up a local node that connects to the Ethereum MainNet during the period of January 23, 2022, to January 25, 2022. When the node received a new block from its neighbor nodes, it performed the block validation process and recorded the three time components. 

\begin{figure}[tp]
	\centering
 \includegraphics[width=\linewidth]{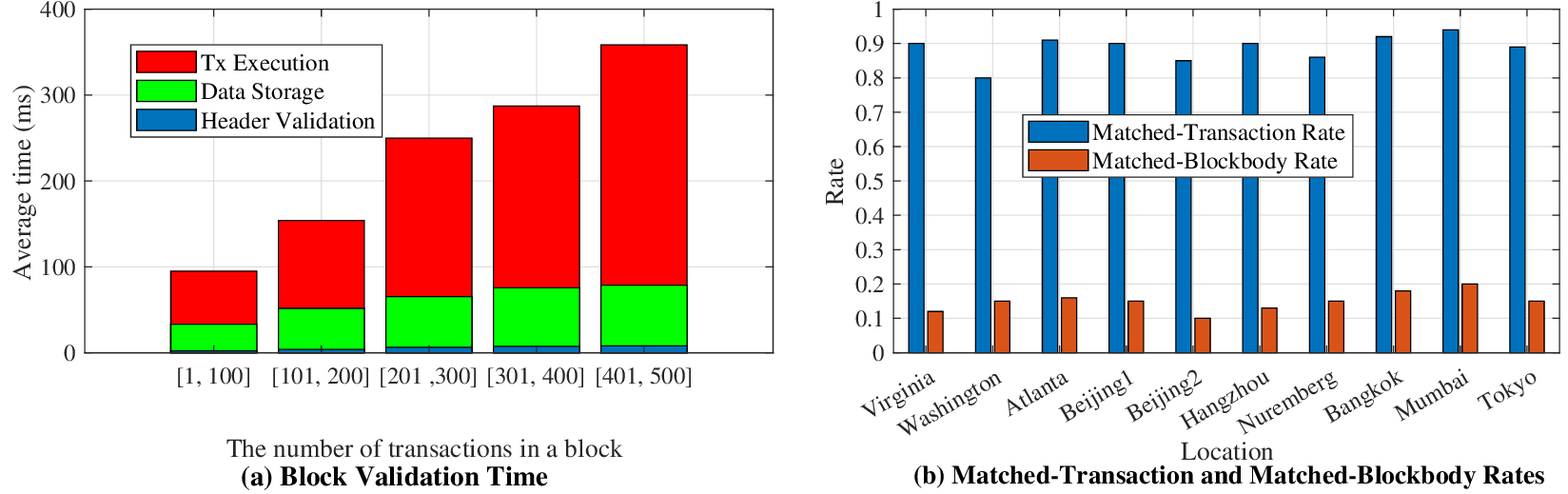}
	\caption{InVestigation Results in Ethereum Blockchain: (a)  Block Validation Time; (b)   Matched-Transaction and Matched-Blockbody Rates.}
	\label{measure1}
\end{figure}

The measured results are summarized in Fig. \ref{measure1}(a). The block validation time and the three components are averaged over blocks containing different numbers of transactions. As observed, the total block validation time increases with the number of transactions in a block, and transaction execution is the component dominating the block validation time. The analysis in \cite{theoreticalBitcoin} shows that block propagation time is mainly determined by block validation time when the block size is large. Thus, the focal point for speeding up block propagation should be to decrease the transaction execution time. 

\subsubsection{Measurement on matched-blockbody probability}\label{matched-block}

As shown in our analysis, the matched-blockbody probability is important for our BBP and other compact-block-like protocols to reduce block transmission time. We measured the matched-blockbody probability in Ethereum. The experimental setup is similar to \cite{hcb}. 
In Ethereum, when a node receives a new transaction, it selects some of its neighbor nodes to forward the new transaction and selects the remaining neighbor nodes to forward the transaction hash \cite{ethna}. If its neighbor nodes receive the transaction hash but the transaction is not in their local transaction pools, the neighbor nodes then request the transaction by sending the transaction hash to the node from which they received the transaction hash. 

In our experiment, we set up a node called ObserverNode. When ObserverNode received a new block from one of its neighbor nodes, it extracted all transactions in the block and converted all these transactions to a set of hashes. It then pretended to miss these transactions and requested them from all its neighbor nodes except the node from which it received the block using the transactions' hashes. Once a neighbor received a transaction request, it responded with the corresponding transaction stored in its local transaction pool (i.e., matched transactions). Once a neighbor responded with all transactions in a block, it matched this blockbody.  




We saved all responses of all neighbor nodes to ObserverNode and measured the matched-blockbody probability. In particular, we measured two statistics. The first is the matched-transaction rate defined as the ratio of the number of matched transactions to the total number of transactions in all recorded blocks. The second is the matched-blockbody rate defined as the ratio of the number of completely matched blockbodies to the number of all recorded blocks. 

The experiment was conducted from December 27, 2021 to January 7, 2022. To ensure good representation and to smooth experimental results, we selected 10 neighbor nodes distributed in different locations around the world. Each of these nodes returned more than 100 responses to  ObserverNode. 
The results, depicted in Fig. \ref{measure1}(b), reveal fluctuating matched-transaction rates between 0.80 and 0.94 among all the neighbor nodes and matched-blockbody rates ranging from 0.10 to 0.20. 
This means that the Ethereum nodes have already stored around 90\% of the transactions of a propagating block into their local transaction pools before they receive the block; but the matched-blockbody probability is only a little more than 10\%  (i.e., there is a good chance that some transactions in a block are still missing).

We can thereby conclude that the matched-blockbody probability in Ethereum is very low (about 10\%) and the compact-block-like protocols do not work in Ethereum. 
By looking deeper into the missing transactions in each block, we found three types of transactions for this low probability: 1) \textbf{Late Transaction} that is broadcast to nodes before the block that contains it is mined by a miner due to network delay; 2) \textbf{Local Transaction} that is selected in the mined block by miners with high priority even though it may be deleted by other nodes due to the low GAS price; and 3) \textbf{Withheld Transaction} that is not broadcast before included in a block due to high profit or privacy, such as the transaction related to FlashBots \cite{flashbots}. More details are referred to in Appendix \ref{trans}.

\subsection{Technique Challenges in BBP}\label{BBPOVERVIEW}

As shown in Fig. \ref{BBPidea}(b), the basic idea of bodyless block propagation (BBP) is that only the blockheader is transmitted between nodes, and a new block is pre-validated so that the in situ block validation during the block propagation is just a simple comparison of the pre-computed global state and the global state embedded in the blockheader.

\begin{figure}[tp]
	\centering
         \includegraphics[width=\linewidth,height=3.8cm]{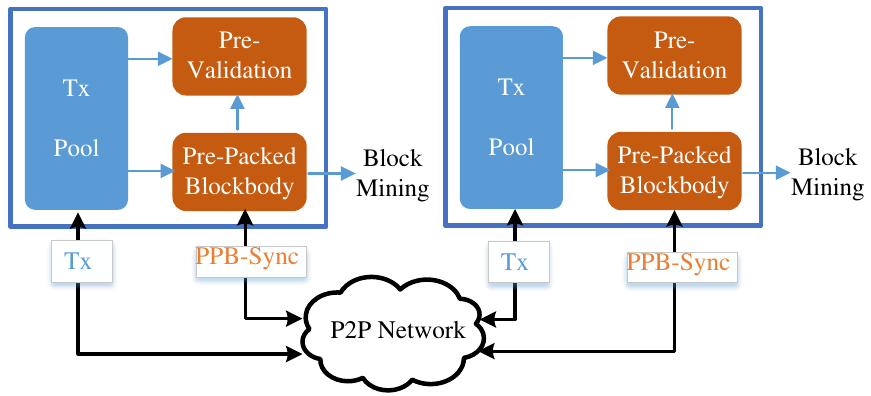}
	\caption{System architecture of BBP by extending TxPool.}
	\label{overFigure}
\end{figure}

To implement BBP, we propose to extend the architecture of Ethereum Tx-Pool implementation by introducing a pre-packed blockbody (PPB) module and a pre-validation module, as depicted in Fig. \ref{overFigure}. The PPB contains a selected subset of transactions from the Tx-Pool, and it is used to generate the next block by the mining module. Each node pre-executes the transactions in its PPB and pre-computes the global state, and then stores the global state in the pre-validation module. 

In the ideal case, the PPBs of all nodes are the same, and this common PPB is the same as the blockbody of the next block mined by the lucky miner. With perfectly synchronized PPBs, only the header of the newly mined block needs to be transmitted. 
When a node receives the new blockheader,  for validation, the node only needs to compare the global state in the blockheader and the global state stored in the pre-validation module. 
However, the ideal scenario may not arise automatically in practice.  We need to overcome several technical challenges for that to happen.

\noindent\textbf{\emph{Challenge 1:}} The first challenge is how to ensure all nodes (or the majority of them) pre-pack the same PPB, even though these nodes are distributed and may behave selfishly. The transactions in Tx-Pool and their packing order may vary from node to node (this is the case for Ethereum). To address this challenge, we propose a Time-specific transaction Selection and Ordering (TSO) algorithm to ensure the ordered transactions in the PPBs of the honest nodes are same with high probability. We further propose a low-overhead PPB synchronization protocol to synchronize the PPBs of neighbor nodes to ensure the PPBs are same in most cases. 

 In blockchain, the miners are selfish and would earn extra profit besides the standard block reward and gas fee, by including, excluding, or reordering the transactions in blocks, which is also called Miner Extractable Value (MEV) \cite{mev_1}. To capture and compete for the MEV opportunity, the miners will listen to the network and may withhold some MEV-related transactions to maximize their benefits, i.e., the withheld transaction in Section \ref{matched-block}. Thus, it is difficult to construct the same block contents (PPB) among miners. In our BBP scheme, this problem is naturally addressed. A mined block different from the PPB in honest nodes (assuming most nodes are honest) has a much larger propagation time since the validation time is longer in the absence of pre-validation information. As a result, such a mined block has a high probability of becoming a stale/uncle block. This disincentivizes the miners from withholding local transactions and disincentivizes the miners from adopting a blockbody different from the ``standard'' PPB of the honest nodes. 

\noindent\textbf{\emph{Challenge 2:}} The second challenge is how to deal with the as-yet unavailable information during the pre-validation process. Even if PPB is perfectly synchronized among all nodes, some required information may not be available yet. For example, some transactions may involve the Coinbase address associated with the miner, but the successful miner is not known yet during the pre-validation process. Our pre-validation algorithm identifies the transaction chains that involve the Coinbase address and only executes transactions that do not involve the Coinbase address. The global state computed during the pre-validation is thus an intermediate global state. Based on the pre-validation result, the re-validation is efficient (if it is needed after the blockheader generated by the successful miner arrives). 

\section{Technique Details of BBP}\label{BBPDesign}
This section presents the detailed design of BBP in Ethereum. We assume the transaction forwarding and PoW consensus protocols of Ethereum as the building components of the blockchain\footnote{Since the implementation aspect of our work is based on Ethereum, we give a brief technical overview of the Ethereum blockchain in Appendix \ref{account}, \ref{select}, and \ref{generate} for readers unfamiliar with the technical details of Ethereum.}. BBP is also compatible with other transaction forwarding and consensus protocols. BBP changes the block forwarding protocol. We give the details of the new elements in BBP: pre-packed blockbody (PPB) generation, pre-packed blockbody pre-validation, pre-packed blockbody synchronization, and bodyless block forwarding protocol. 

\subsection{PPB Generation}\label{PPB_generation}


In BBP, each node retrieves a subset of transactions from its local Tx-Pool in a specific order to generate a PPB. In the ideal case, all nodes have the same PPB and use the same PPB for pre-validation. In Ethereum, however, the transaction selection and ordering are decided individually by the miners, with preference given to their own transactions and those with high GAS prices. To promote a common PPB across nodes, we propose a Time-specific transaction Selection and Ordering (TSO) algorithm to disambiguate transactions and their order.


\noindent\textbf{\emph{TSO:}} As discussed in Section \ref{matched-block}, late transactions and local transactions are two causes for dissimilar PPB at different nodes. To eliminate these two causes, TSO follows three new rules as described in \emph{Algorithm \ref{alg1}}. First, only transactions whose timestamp is earlier than a predetermined timestamp threshold $T$ are eligible to be selected. Second, among all the eligible transactions, transactions with a higher GAS price are selected and ordered with a higher priority. Finally, if two or more selected transactions have the same GAS price, their order is determined by their transaction hashes. The detailed TSO algorithm is given in \emph{Algorithm \ref{alg1}}\footnote{ In \emph{Algorithm \ref{alg1}}, we need to select the transaction with the smallest nonce for each account, as specified by line 1 and line 11, i.e., the first transaction.}.

	
		
		

In Ethereum, even if different nodes select the same set of transactions into one block, the transaction orders may be different because varying network latencies may cause different timestamps for the same transaction at different nodes. In TSO, the order is unambiguous once different nodes select the same set of transactions through the timestamp threshold $T$ (once selected, GAS price and TxHash dictate the order of the transactions, not the timestamp). However, varying network latencies may still cause a problem for transaction selection, where the selected transactions with a timestamp around $T$ may be slightly different among nodes. We propose a two-pronged approach to address this issue: time threshold setting and PPB synchronization (see Section \ref{PPB-SP}  for details). 

\noindent\textbf{\emph{Time Threshold Setting:}} As mentioned above, each node only picks the transactions received earlier than $T$. Our first remark on $T$ is that all nodes can easily agree on the same timestamp $T$ for each block. The second point is that a small value of $T$ may help to guarantee that the transactions to be selected will have arrived at all nodes when pre-validation is to be done, but the selection of a small $T$ may cause a larger transaction commit latency. We propose to set $T$ as the timestamp carried in the current block, which is known by all the nodes that have received the current block. This way, the commit latency is also acceptable.
\begin{algorithm}[t]
	\caption{TSO algorithm} 
	\label{alg1}
	
	\begin{algorithmic}[1]
\footnotesize
		\REQUIRE a timestamp threshold $T$; two empty sets $G$ and $G_c$; the transaction queue $I[i]$ for each account $i$, in which the transactions of account $i$ are sorted in ascending order according their nonces. 
		\\    
		\ENSURE $G$	
		\STATE Copy the first transactions of the all transaction queues $I[i]$ to the candidate set $G_c$.
		\STATE Remove the transactions in $G_c$ with timestamp later than $T$. 
		\WHILE{The block GAS limit is not reached} 
		\STATE  Select the transaction in $G_c$ with the highest GAS price.
		\IF{More than one transaction have the same highest GAS price} 
		\STATE Select the transaction in $G_c$ with the highest GAS price and the highest Hash value.
		\ENDIF
		\STATE Append the selected transaction to $G$ and remove it from $G_c$.
		\STATE Remove the selected transaction from its transaction queue $I[k]$.
		\IF{The first transaction of the transaction queue $I[k]$ has a timestamp earlier than $T$} 
		\STATE Copy this transaction to $G_c$.
		\ENDIF
		\ENDWHILE
		
	\end{algorithmic}
\end{algorithm}
\subsection{Pre-Validation of PPB}



With the help of TSO and the PPB synchronization scheme (Section \ref{PPB-SP}), BBP guarantees that the majority of the nodes have the same PPB for the pre-validation process. Our pre-validation module computes tentative validation information consisting of three components. The first component $BodyHash$ is similar to the field of $TXs$-$Hash$ in the Ethereum block header, which is computed from all the transactions in PPB. The second component \emph{UnexecutableTxs} is all the transactions that cannot be executed in advance. The last one is the \emph{IntermediateState}. We explain the second and last component below:

\noindent\textbf{\emph{Un-executable Transactions:}} In Ethereum and most emerging blockchains supporting Turing-complete computing tasks with smart contracts, some transactions may involve the Coinbase address and cannot be executed in advance. One example is the MEV transaction that may invoke an instruction of smart contract to privately pay tips to the Coinbase address so that it can be sealed into the block as soon as possible \cite{flashbots}. But for PPB pre-validation, the Coinbase address is not yet known, and therefore Coinbase transactions cannot be executed in advance. Similarly, all other transactions affected by the Coinbase transactions cannot be executed in advance. For PPB, we need to first identify all the un-executable transactions among the selected transactions.

\begin{figure}[tp]
	\centering
        \includegraphics[width=\linewidth]{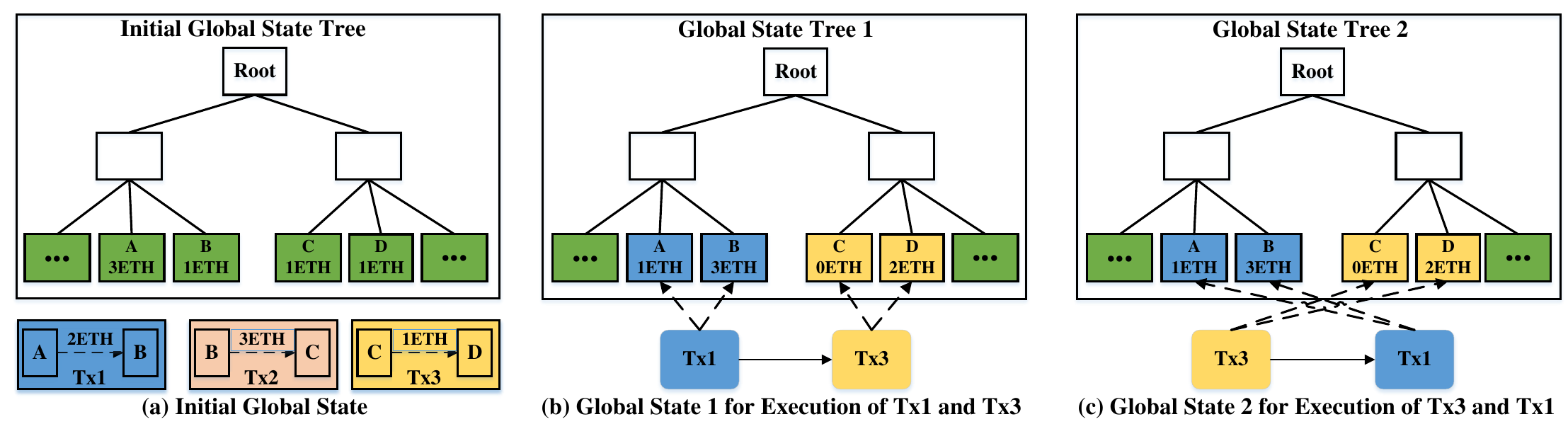}
	\caption{Execution for un-intersected transactions.}
 \vspace{-10pt}
	\label{intersect}
\end{figure}

We say that two transactions intersect if there is at least one common accessed account between them. For example, as shown in Fig. \ref{intersect}(a), the accessed accounts of Tx1 are \{A, B\} (Tx1: transfer 2 ether from account A to account B), the accessed accounts of Tx2 are \{B, C\}, and the accessed accounts of Tx3 are \{C, D\}. Tx1 and Tx2 intersect with the common account \{B\}, and Tx1 and Tx3 do not intersect. In general, if two transactions intersect, the global state depends on their execution order (e.g.,  if both Tx1 and Tx2 are packed in one block, to obtain the same global state, they should be executed in order). If two transactions do not intersect, they can be executed in any order without affecting the final global state (e.g., as shown in Fig. \ref{intersect}(b) and (c), if both Tx1 and Tx3 are packed in one block, they can be executed successfully in any order).



Based on this definition, we can obtain the sequence of all un-executable transactions, sequence $U_g$, from PPB.  We first use Ethereum Virtual Machine (EVM) to find the transactions that invoke the related instruction to access the Coinbase address and append them to $U_g$. Then, all the transactions that intersect with any transaction in $U_g$ are append to $U_g$ in order. This process is repeated until all remaining transactions in PPB do not intersect with any transaction in $U_g$.


\noindent\textbf{\emph{Intermediate State:}} The transactions in PPB that are not in $U_g$ can be executed independently of the transactions in $U_g$. Thus, we can execute the transactions not in $U_g$ before the next block is mined. Specifically, we first execute the remaining transactions in PPB not in $U_g$ sequentially in the same order as in PPB and store the resulting global state as the intermediate state. The detailed pre-validation algorithm is shown in \emph{Algorithm \ref{alg2}}.


\subsection{PPB Synchronization Protocol}\label{PPB-SP}



As outlined in Section \ref{PPB_generation}, PPBs at different nodes may still differ due to transactions with a timestamp near $T$.  We propose a simple and secure PPB synchronization protocol with which neighbors exchange information to align their PPBs\footnote{Notably, PPB synchronization protocol also needs additional communication. If the nodes reach a consensus about PPB, they need only propagate the block header and use the pre-validation results of PPB to speed up block propagation significantly.}.


In the PPB synchronization, Node 1, when assembling or modifying its PPB, sends a $checkSync$ message containing the $BodyHash$ to its neighbors, such as Node 2. If Node 2 detects the received $BodyHash$ is different from its own, it sends back all the transactions in its local PPB to Node 1. In this way, Node 1 and Node 2 can locate those different transactions\footnote{The transaction localization may be improved with the algorithms in \cite{graphene}.}. Simply combining the eligible transactions, Node 1 and Node 2 add them to both PPBs sorted with TSO. In this way, we can finally obtain a synchronized PPB in theory, which is the union of all the nodes' initial PPBs. 


To deal with the potential of a dishonest node's attack by inserting ineligible transactions in PPB, we modify the above protocol a little. When a node finds some transactions in its neighbor's PPB that are not in its own PPB, only the transactions in its local Tx-pool with local timestamps earlier than $T+\Delta$ are added to its own PPB. For transactions not in its local Tx-pool or transactions with a local timestamp later than $T+\Delta$, the node cannot add them to its local PPB. Therefore, if an attacking node purposely modifies the timestamp of a transaction to be earlier than the threshold $T$, an honest node will reject this invalid transaction since the local timestamp at the honest node is later than $T+\Delta$. Here, $\Delta$ is a predetermined parameter to upper bound the transaction propagation delay. The work in \cite{ethna} measured that a transaction would take no more than 1 second to propagate to most standard Ethereum nodes, and thus we set $\Delta=1$ second.

\begin{algorithm}[t]
	\caption{Pre-validation algorithm} 
	\label{alg2}
	\begin{algorithmic}[1]
\footnotesize
		\REQUIRE ~initial state $S_0$; the remaining transaction sequence  $Tx=[Tx1,Tx2,...,Txn]$\\                  
		\ENSURE $S_{j+1}$
		\STATE $i=0$, $j = 0$
		
		\WHILE{$i<n$} 
		\IF{$S_{j+1}= Exec(S_j,Tx[i])$ is successful} 
		\STATE $j=j+1$
		\ELSE 
		\STATE Remove $Tx[i]$ from PPB
		\ENDIF
		\STATE $i=i+1$
		\ENDWHILE
	\end{algorithmic}
\end{algorithm}

\subsection{BBP Forwarding Protocol}

When a miner successfully mines a block, it sends the blockheader to its neighbors with the same $BodyHash$ and the full block to its neighbors with different $BodyHash$. When a node receives the blockheader, it conducts the PPB-validation procedure detailed in the next paragraph. If validation passes, the header is forwarded. In rare cases of different PPBs among a few nodes due to imperfect PPB synchronization, the miner sends the full block to a neighbor with a different $BodyHash$\footnote{The neighbors' $BodyHash$ values can be remembered during the PPB synchronization procedure. This broadcasting can be more efficient by attaching those missing transactions in the blockheader for the rare cases of non-synchronized PPBs.}.

\noindent\textbf{\emph{PPB-Validation Procedure:}} With the pre-validation information, a node can quickly verify the received blockheader. First, it compares the local $BodyHash$ with the $Txs$-$Hash$ in the received blockheader. If the two hashes are different, the blockbodies are different, and the node requests the full block for validation as in Ethereum. If the $BodyHash$ check passes, the node then gets the Coinbase address from the blockheader, with which the node can then execute the transactions in $U_g$ in order, based on the $IntermediateState$. If the final global state equals the one in the blockheader, this block will be committed to the local database, and the validation is successful.  
\begin{figure}[tp]
	\centering
 \includegraphics[width=\linewidth,height=2.3cm]{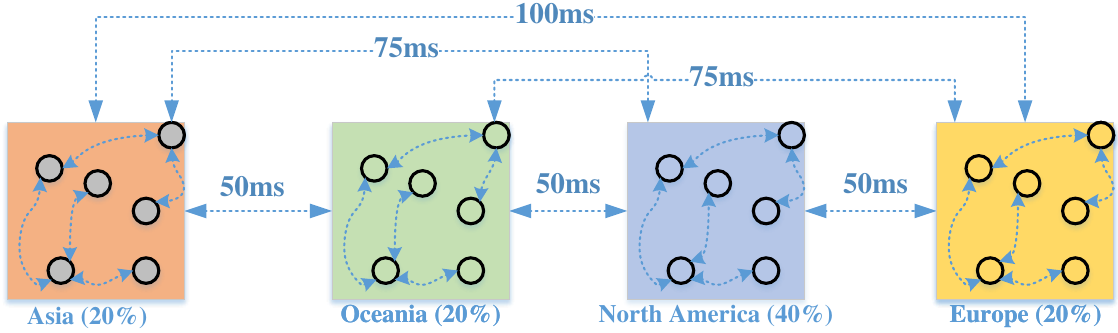}
	\caption{The nodes distribution and expected link delays in our experiments.}
	\label{Node}
\end{figure}
\section{Experimental Evaluation and Performance Discussion}\label{expeva}
In this section, we implement and evaluate our BBP scheme over a test network. For comparison, we also implement three typical block propagation protocols: the Legacy Block Propagation (LBP) and Compact Block Propagation (CBP) of Bitcoin, and the Block-Hash Propagation (BHP) of Ethereum. 


\subsection{Typical Block Propagation Protocols}
LBP, CBP, and BHP are the mainstream block propagation protocols in the public blockchain systems, and their specific process can be referred to \cite{theoreticalBitcoin,compact,ethna}. 

\textbf{LBP} is the original block propagation protocol used in Bitcoin. When a node receives a block, it first validates the blockheader and blockbody. If the block passes validation, it announces an Inv message to its neighbor nodes. If a neighbor node did not receive this block before, it replies with a GetData message to request the full block. 
 
\textbf{CBP} is the current block propagation protocol adopted in Bitcoin to further reduce the network overload. When a node receives a new block, it validates the block and generates a compact block version. Then it announces this compact block by sending an Inv message to its neighbor nodes. If a neighbor node does not receive the compact block, it only needs to request this compact block and some transactions missed in its local transaction pool.

\textbf{BHP} is a hybrid block propagation protocol adopted in Ethereum. When a node receives a  new block, it selects $\sqrt X$  neighbor nodes to forward the full block after it simply verifies the blockheader, where $X$ is the number of neighbor nodes that have not received the new block. Then, the node further verifies the full block and forwards the block hash to the remaining $(X- \sqrt X)$ neighbor nodes. For the neighbor nodes that received the block hash, they will wait for a certain period to fetch the blockheader and blockbody in turn.

\subsection{Experiment Setup}

\textbf{\emph{Experiment Testbed:}} The direct deployment of thousands of physical machines over a large blockchain network topology is complicated and expensive, if not impossible. To run our experiments, we built a multi-node network environment based on docker container technology \cite{docker} in the Linux server and physically connected 4 servers over a software-defined network. Each server deploys multiple docker containers and allocates its hard resources (e.g., CPU, RAM, and SSD) to them. Each docker container is labeled as a physical machine located somewhere in the world, and the propagation latency between two containers is configured by the geographical label. With this software-configurable semi-physical testbed, we can simulate large-scale distributed networks. We fully implemented the blockchain peers for the above four block-propagation protocols based on the Ethereum software version v1.8 \cite{GoEthereum}. The blockchain peer software was loaded and run in each docker container. Performance data are recorded in a log file and analyzed with shell/python programs.

\renewcommand{\arraystretch}{0.9}
\begin{table}[]\centering
  \footnotesize
	\caption{Parameter of the Docker network simulator}
      \vspace{-5pt} 
	\begin{tabular}{ll}
	    \toprule
		Parameter                                                                  & Values                                                                                     \\
		\midrule 
		Number of nodes                                                            & 1000                                                                                       \\ 
		Block Interval                                                             & 13-15s \cite{etherscan}                                                                                      \\ 
		Network topology                                                           & \begin{tabular}[c]{@{}l@{}}Power-law distribution and \\ small-world property \cite{ethna} \end{tabular} \\ 
		\begin{tabular}[c]{@{}l@{}}Average bandwidth \end{tabular} & 55Mbps \cite{decentralization}                                                                                    \\ 
		Packet loss rate for a link                                             & Randomly distributed over (0\%, 1\%)                                                          \\ 
		\bottomrule
	\end{tabular}
\label{network}
\end{table}

\begin{figure*}[tp]
	\centering
        \includegraphics[width=\linewidth]{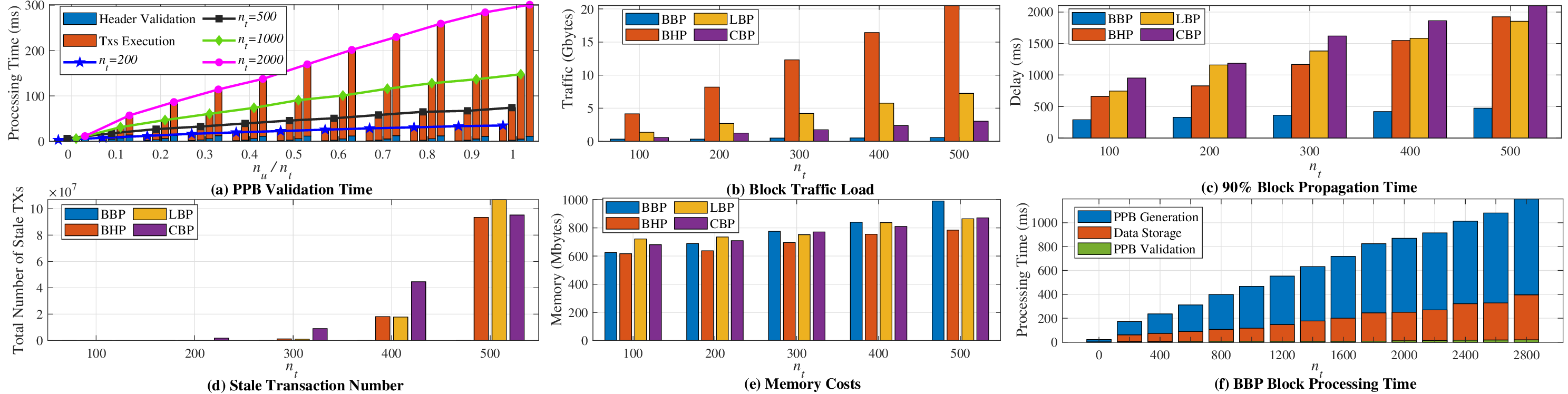}
 \caption{Basic BBP Performance: (a) PPB validation time versus number of transactions $n_t$  number of non-executable transactions $n_u$;  (b) Network traffic generated by block propagation versus blocks with various transaction number $n_t$ in a block; (c) 90\% block propagation time with various $n_t$; (d) Number of stale transactions within a period of 1,000 blocks; (e) Memory Costs with various $n_t$; (f) BBP block processing time with various $n_t$. }
	\label{basic1}
\end{figure*}

\noindent\textbf{\emph{Network Parameters and Topology:}} 
The critical aspect of our experimental testbed is the network environment and topology. We developed a network controller capable of configuring and managing network parameters between any two nodes in the testbed. Following the worldwide nodes' geographical distribution as in \cite{ethernodes}, we categorized experimental nodes into four groups representing dominant continents to simulate global node distribution: Asia, Oceania, North America, and Europe. Utilizing ping-measured link propagation latency, we established link latency distributions. Intra-group links exhibit a uniform distribution over [10ms, 40ms] with a mean of 25ms. Inter-group links follow varying uniform distributions: [40ms, 60ms] with a mean of 50ms, [60ms, 90ms] with a mean of 75ms, and [90ms, 110ms] with a mean of 100ms. The detailed mean delays of inter-group links are illustrated in Fig. \ref{Node}. Additionally, the blockchain's overlay P2P network is generated using the Ethereum node discovery algorithm, mirroring Ethereum's approach.

\noindent\textbf{\emph{Experimental Testbed Validation:}} In our experiment, we built a testbed network to simulate  Ethereum with parameter settings as in Table \ref{network}. As noted in Table \ref{network}, most parameter settings are drawn from prior work that studied Ethereum.

To validate our testbed, we compare the 90\% block propagation delay of our testbed with real Ethereum. We calculated the uncle block probability $p_f$ (i.e., fork probability) in Ethereum based on the number of daily uncle blocks \cite{etherscan} in four periods respectively: 1) 21/12/2021-31/12/2021; 2) 17/02/2022-26/02/2022; 3) 15/03/2022-24/03/2022; and 4) 11/04/2022-20/04/2022.
According to the measurement result, $p_f$ fluctuated within the range of [0.053, 0.067] with an average value of 0.059.  Additionally, the block interval is 13-15s during this period, with an average value of 14s. Using eq. (\ref{fork}), we compute that the 90\% block propagation delay on the actual Ethereum network is about 851ms. For our testbed experimental results, as shown in Fig. \ref{basic1}(c), when $n_t$  is 200,  the 90\% block propagation delay of Ethereum protocol is 829ms, close to the delay on the actual network. Thus, our testbed is a good tool to simulate the real Ethereum network. 


\subsection{Experiment Results} \label{experiment results}
This subsection presents experimental results about BBP performance in two parts: 1) Basic BBP performance under the network without any malicious nodes; and 2) BBP robustness to defend against various attacks.

\subsubsection{Basic BBP performance}\label{Basic_BBP_Performance}

This part presents experimental results to validate the efficiency of BBP by comparing its basic performance with LBP, BHP, and CBP under the testbed without any malicious nodes. Specifically, we examine block processing time, block traffic load, memory costs, and block propagation delays of these propagation protocols. 

In this testbed, simple transfer transactions are considered, and the transaction generation rate is adjusted according to the practical transaction generation rate (24 per second) and the practical number of transactions in a block (200) in Ethereum MainNet \cite{etherscan}.  For example, when the number of transactions in a block in the testbed is 400, the transaction generation rate is set as $\frac{400}{200}*24=48$ per second.   

\noindent\textbf{\emph{PPB Validation Time: }} To demonstrate the efficiency of the pre-validation algorithm, we measure the PPB validation time when a BBP node receives a blockheader, including the validation of the blockheader and execution of transactions in the non-executable sequence. The experiment results are shown in Fig. \ref{basic1}(a), with different block transactions $n_t$  and different sizes of the non-executable sequence $n_u$. From Fig. \ref{basic1}(a), we can see that the validation time is minimized when $n_u=0$  and is about  12ms for $n_t=2000$.  As $n_u$ increases, the PPB validation time increases linearly. Based on Ethereum data \cite{etherscan}, each block usually contains 1-3 Coinbase transactions, i.e., $n_u$ is around 1\% to ensure efficient PPB validation.


\noindent\textbf{\emph{Block Traffic Load:}} We then measure the total network traffic induced by broadcasting blocks among all nodes. The measurement results are shown in Fig. \ref{basic1}(b). Benefiting from the bodyless design, the block traffic load of BBP almost keeps constant for various $n_t$, while the loads of other three protocols increase linearly with $n_t$. When $n_t=200$, the block traffic of  BBP is only 1/2 that of CBP, 1/4 that of LBP, and 1/10 that of BHP.  BHP creates the most traffic load because it broadcasts full blocks to $\sqrt X$ neighbors, while CBP and LBP only request one full block when needed. It should be noted that LBP and CBP incur one more round of communication.

\noindent\textbf{\emph{Block Propagation Time:}} Fig. \ref{basic1}(c) shows the time for a block to propagate to 90\% of nodes in the network versus the number of transactions in a block ($n_t$). Due to the scalability of BBP, its propagation time is almost constant for various $n_t$, while for other protocols, the propagation time linearly increases with $n_t$. Specifically, BBP needs the least amount of time among all protocols, and it only needs 1/4 time of the BHP protocol when $n_t=500$. The CBP performs worst in the setting since the block interval $t_g$ (13-15s) is much shorter than that (10 minutes) in Bitcoin. As with our measurement in Section \ref{matched-block}, over 90\% nodes need to request the missing transactions with one more round of communications.

\noindent\textbf{\emph{Stale Transaction Number:}} When $n_t$ increases, not only the block propagation time increases but also the number of stale transactions. A received transaction is regarded as a stale transaction if it has been committed into a block according to local recording. In this case, the transaction propagation lags the block propagation, and the blockchain network does not work properly anymore. As shown in Fig. \ref{basic1}(d), the number of stale transactions for BBP is near zero while the numbers of stale transactions for other block propagation protocols increase exponentially when $n_t > 300$. The reason is that BBP only packs transactions prior to the current block and leaves more time for the transactions to propagate.

\noindent\textbf{\emph{Memory Cost:}} While our BBP achieves a significant benefit on block propagation, additional memory cost is required, since each BBP node needs to generate and store PPB before the next block arrives. We measured the detailed memory costs for four block propagation protocols. The experimental results are shown in Fig. \ref{basic1}(e). As observed, with the increase of $n_t$, memory costs for BBP increase fastest among the four protocols, and BBP always requires a larger memory than BHP. Especially BBP requires around 1.3x memory than BHP when $n_t$ is 500. Given the significant advantage of achieving low propagation delay with BBP as shown in Fig. \ref{basic1}(c) and the standard memory requirement (32Gbytes) suggested by Ethereum \cite{GoEthereum}, this additional memory cost is acceptable.

\noindent\textbf{\emph{TPS Limited by Node:}} While BBP holds the potential for nearly constant block propagation time, its achievable TPS is mainly constrained by the BBP block processing time at each node, i.e., the EVM limitation. For an ideal case, BBP can finish the network consensus as well as a new block is successfully processed by the nodes. To ensure the maximum TPS of BBP, we measure the specific BBP block processing time for one node in Fig. \ref{basic1}(f), including PPB generation, PPB validation, and data storage. As observed, the BBP block processing time increases linearly with $n_t$ and exceeds 1 second when $n_t> 2,400$. In other words, 2400 is the maximum TPS for one node without propagation time. Thanks to the almost constant and small block propagation time achieved by BBP, we can obtain a TPS of 2335, near the maximum TPS, by adjusting $n_t\approx18,857$ and block interval $t_g\approx 8$ seconds. More details to obtain $n_t$ and $t_g$ can be referred to in Appendix \ref{maximum_tps}.   



\subsubsection{BBP robustness}

BBP accelerates block propagation by generating consistent PPB among different nodes, proving advantageous in a network without malicious nodes, as demonstrated in Section \ref{Basic_BBP_Performance}. However, in the practical blockchain network, three types of potential attacks may interfere with the PPB generation to offset the benefit: 1) \textbf{Attack I} that malicious nodes delay or do not broadcast transactions to prevent the honest nodes from generating the correct PPB; 2) \textbf{Attack II} that malicious nodes generate invalid PPBs to infect honest nodes by PPB synchronization; and 3) \textbf{Attack III} that malicious nodes delay or do not broadcast transactions while generating invalid PPBs.

To validate the BBP robustness in the network with these attacks, we measure the proportion of non-synchronized PPBs for BBP and compare 90\% block propagation times for BBP and BHP under the testbed network with various malicious nodes. Note that 90\% block propagation time for BHP is only counted under the network with malicious nodes issuing Attack I since Attack II and III are only related to PPB generation. In this testbed network, we replay transactions from block numbers 11,503,300 to 11,548,969 in Ethereum MainNet, aligning with the range examined in \cite{mev_1}, which analyzes transactions that may involve Coinbase Address. Also, both the transaction generation rate and the average number of transactions in a block mirror those observed in the actual Ethereum MainNet \cite{etherscan}.


\begin{figure}[tp]
	\centering
        \includegraphics[width=\linewidth]{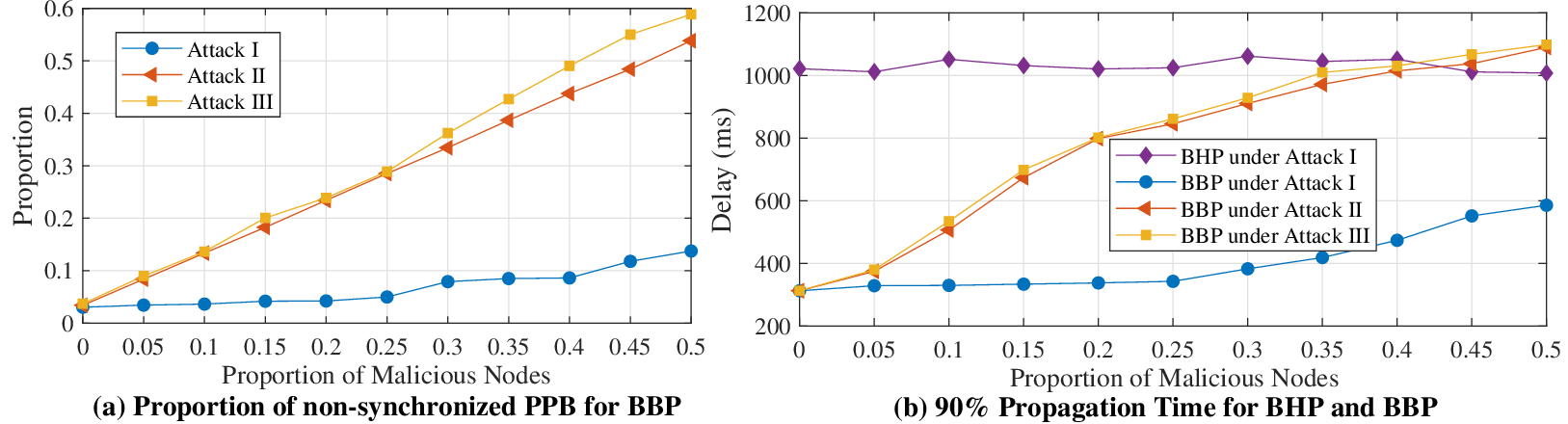}
	\caption{BBP Robustness for the network with various malicious nodes: (a) Proportion of Non-Synchronized PPB; (b) 90\% Block Propagation Time. }

	\label{attack}
\end{figure}

The experimental results are depicted in Fig. \ref{attack}: (a) the non-synchronized PPBs and (b) 90\% Block Propagation Time. From Fig. \ref{attack}, it is evident that the average proportion of non-synchronized PPBs, in the absence of malicious nodes, is approximately 3.5\%. Since 90\% node propagation is good enough, this 96.5\% proportion of synchronized PPBs is a good result. Also, BBP exhibits robustness under Attack I when the proportion of malicious nodes is below 0.25. In this scenario, the percentage of non-synchronized PPBs consistently stays below 6\%, ensuring the attainment of a 90\% block propagation time of less than 350ms. This underscores the efficacy of BBP in effectively mitigating Attack I. However, BBP demonstrates weak robustness against Attacks II and III, as both the proportion of non-synchronized PPBs and 90\% block propagation time escalate rapidly with the increasing presence of malicious nodes. Especially when the percentage of malicious nodes issuing Attack II or III exceeds 40\%, the 90\% block propagation time of BBP is larger than that of BHP. Thus, in practical scenarios, Attacks II and III would introduce challenges to the seamless implementation of BBP.

\noindent\textbf{\emph{Countermeasure for Attacks II and III: }} Attacks II and III prolong the block propagation delay of BBP by violating TSO algorithms to generate invalid PPBs, e.g., selecting transactions with low GAS prices rather than high GAS prices. Fortunately, malicious nodes with attacks II and III can be simply identified by a score function implemented in GossipSub \cite{gossipsub}. That is, each BBP node keeps a score for every neighbor node and adds/subtracts this score according to the validity of the PPBs received by the corresponding neighbor node. The BBP node would disconnect the malicious neighbors whose score is lower than a threshold.


\section{Related Work}\label{RELATEWORK}
Limited TPS is a fundamental problem for public blockchain, and there are quite a few works on improving TPS, such as DAG, Sharding, new consensus protocols, and layer 2 \cite{14Sec, 15aaa}. DAG is a new data structure that all transactions are directly or indirectly connected and can be validated in parallel \cite{iota, byteball}; Sharding is a widely applied solution by dividing the nodes and related transactions into different groups for parallel processing \cite{ELASTICO, dang2019towards, rapidchain, 23aa, polkadot}; 
the new consensus protocols, such as Proof-of-Stake \cite{casper}, Practical Byzantine Fault Tolerance \cite{25aa}, HoneyBadgerBFT \cite{honeyBadgerBFT}, Dumbo \cite{dumbo}, Hotstuff \cite{Hotstuff}, upgrade PoW adopted by Bitcoin and Ethereum 1.0 \cite{pow_1} to directly speed up block generation and TPS; layer 2 is an off-chain solution that uses separate chains to reduce the number of transactions executed on the main chain, such as Plasma \cite{17plasma}, Optimistic Rollup \cite{optimisticrollup,optimistic_1}, and ZK-Rollup \cite{zkrollup}.


Speeding up block propagation is a more primary way to improve the blockchain TPS performance since the blockchain is built on the network-wide broadcast of blocks \cite{a2}. More importantly, as a networking functionality, these schemes are well-suited for practical systems and compatible with diverse consensus algorithms and other TPS-scaling strategies. As discussed in Section \ref{introduction}, speeding up block propagation falls into three categories: 1) Compressing block sizes; 2) Simplifying block validation; 3) Optimizing network topologies.


 \noindent\textbf{\emph{Compressing block sizes:}} One way to speed up block propagation is compressing the block size to shorten the block transmission time. Compact block is first adopted in Bitcoin \cite{compact}, where the complete transactions in the blockbody are replaced with their hash, and the receiver reconstructs the full block based on its transaction pool. As an enhancement of compact block, Xthin block protocol proposed to send all transaction information in the receiver's transaction pool (as a bloom filter) to the sender to piggyback the missing transactions in the compact block \cite{xthinblock}. Graphene combines Bloom Filters and Invertible Bloom Lookup Tables to further compress the block size and quickly reconstruct the full block at the receiver \cite{graphene}. 
Txilm ensures optimal transaction hash size by assessing collision probability and incorporating a "salt" during hash calculation to safeguard against potential attacks \cite{txilm}.  The improvement of simply compressing the blockbody is bounded, Dino protocol alternatively transmits a block reconstruction rule to reduce the bandwidth consumption and shorten the block transmission time \cite{dino}.
 

Following the idea of compressing the block size, several works introduce network coding to multicast the blocks \cite{a3, code}. Velocity utilizes fountain codes to enable the nodes to receive a full block from multiple neighbor nodes \cite{a3}. The authors in \cite{code} design a new compact block protocol with cut-through forwarding and incorporate fountain codes to use the upload bandwidths of all neighbor nodes efficiently. 

\noindent\textbf{\emph{Simplifying block validation:}} It is shown that block validation is also a bottleneck in block propagation for large block size \cite{theoreticalBitcoin}, and several works \cite{reducingfork,Heco,Sui} deal with this problem. The work \cite{reducingfork} proposed to probabilistically validate received new blocks rather than to validate all blocks. Some works, e.g. \cite{Heco, Sui}, decouple the process of block validation and consensus and further validate the unrelated transactions in the block in parallel to improve the validation efficiency. 

\noindent\textbf{\emph{Optimizing network topologies:}} Blockchain nodes usually form a complex network and optimizing the network topology can also shorten the block transmission time. A structured network topology that a node broadcast blocks to all nodes efficiently was proposed in \cite{kadcast}. Urocissa \cite{urocissa} maintains multiple broadcasting trees in a distributed way to speed up block propagation and enhance system stability. FRing \cite{fring} emphasizes geography-based topology for lower redundancy and latency. Structured networks, while efficient, entail maintenance overhead and weaken robustness compared to unstructured ones. Several works \cite{Fiber, Falcon, bloxroute} also advocate embedding extra relay networks, which requires new network infrastructure controlled by a central authority.



At first glance, our BBP scheme appears to combine compact blocks and simplified validation. However, BBP fundamentally alters the block transmission and validation workflow by introducing the concept of pre-packed blockbody. As a result, BBP is the first scheme to completely remove blockbody transmission (ignore retransmission probability) and block validation time. Thus, we call it truly scalable since each block duration is independent of the transaction volume in the block.  Also, BBP achieves such significant improvement without any sacrifice of security or decentralization. 


\section{Conclusion}\label{CONCLUSION}

The block propagation time in current blockchain networks, consisting of block validation time and transmission time, is the TPS performance bottleneck. Furthermore, the tradeoff between TPS and security is fundamental in today's blockchains: many solutions that boost TPS come at the expense of lowered security. This paper proposes a bodyless block propagation (BBP) mechanism with pre-validation that can speed up block propagation by 4 times relative to Ethereum without compromising security. More importantly, the block propagation time in BBP is almost independent of the number of transactions in a block. As a result, block propagation no longer limits TPS. Implementing BBP amounts to just inserting a pre-packed blockbody module and a pre-validation module as extensions to the transaction pool. Therefore, BBP is compatible with other lower-layer and upper-layer blockchain techniques. Both the theoretical analysis and experimental results corroborate the TPS scalability and superior performance of BBP.

\clearpage

\bibliographystyle{IEEEtran}

\section*{Acknowledgement}
The research is partially supported by research grants from the Chinese NSF project (62171291) and Guangdong Basic and Applied Basic Research Foundation (2019B1515130003).

\bibliography{reference}

\begin{thebibliography}{10}
\providecommand{\url}[1]{#1}
\csname url@samestyle\endcsname
\providecommand{\newblock}{\relax}
\providecommand{\bibinfo}[2]{#2}
\providecommand{\BIBentrySTDinterwordspacing}{\spaceskip=0pt\relax}
\providecommand{\BIBentryALTinterwordstretchfactor}{4}
\providecommand{\BIBentryALTinterwordspacing}{\spaceskip=\fontdimen2\font plus
\BIBentryALTinterwordstretchfactor\fontdimen3\font minus
  \fontdimen4\font\relax}
\providecommand{\BIBforeignlanguage}[2]{{%
\expandafter\ifx\csname l@#1\endcsname\relax
\typeout{** WARNING: IEEEtran.bst: No hyphenation pattern has been}%
\typeout{** loaded for the language `#1'. Using the pattern for}%
\typeout{** the default language instead.}%
\else
\language=\csname l@#1\endcsname
\fi
#2}}
\providecommand{\BIBdecl}{\relax}
\BIBdecl

\bibitem{bitcoin}
S.~Nakamoto, ``Bitcoin: A peer-to-peer electronic cash system,''
  \emph{[Online]. Available: http://bitcoin.org}, 2008.

\bibitem{ethereum}
V.~Buterin \emph{et~al.}, ``Ethereum: A next-generation smart contract and
  decentralized application platform,'' \emph{URL https://github.
  com/ethereum/wiki/wiki/\% 5BEnglish\% 5D-White-Paper}, vol.~7, 2014.

\bibitem{wood2014ethereum}
G.~Wood \emph{et~al.}, ``Ethereum: A secure decentralised generalised
  transaction ledger,'' \emph{Ethereum project yellow paper}, vol. 151, no.
  2014, pp. 1--32, 2014.

\bibitem{mev_1}
K.~Qin, L.~Zhou, and A.~Gervais, ``Quantifying blockchain extractable value:
  How dark is the forest?'' in \emph{2022 IEEE Symposium on Security and
  Privacy (SP)}, 2022, pp. 198--214.

\bibitem{NFT_2}
L.~Ante, ``Non-fungible token (nft) markets on the ethereum blockchain:
  temporal development, cointegration and interrelations,'' \emph{Economics of
  Innovation and New Technology}, vol.~32, no.~8, pp. 1216--1234, 2023.

\bibitem{metaverse}
H.~Duan, J.~Li, S.~Fan, Z.~Lin, X.~Wu, and W.~Cai, ``Metaverse for social good:
  A university campus prototype,'' in \emph{Proceedings of the 29th ACM
  International Conference on Multimedia}, 2021, pp. 153--161.

\bibitem{ethereum_tps}
M.~Sch{\"a}ffer, M.~Di~Angelo, and G.~Salzer, ``Performance and scalability of
  private ethereum blockchains,'' in \emph{Business Process Management:
  Blockchain and Central and Eastern Europe Forum: BPM 2019 Blockchain and CEE
  Forum, Vienna, Austria, September 1--6, 2019, Proceedings 17}.\hskip 1em plus
  0.5em minus 0.4em\relax Springer, 2019, pp. 103--118.

\bibitem{bitcoinTPS}
E.~Georgiadis, ``How many transactions per second can bitcoin really handle?
  theoretically.'' \emph{Cryptology ePrint Archive}, 2019.

\bibitem{etherscan}
``Etherscan,'' \url{https://etherscan.io/txsPending/}, 2015, accessed January,
  2022.

\bibitem{blocksecurity}
A.~Gervais, G.~O. Karame, K.~W{\"u}st, V.~Glykantzis, H.~Ritzdorf, and
  S.~Capkun, ``On the security and performance of proof of work blockchains,''
  in \emph{Proceedings of the 2016 ACM SIGSAC conference on computer and
  communications security}, 2016, pp. 3--16.

\bibitem{compact}
M.~Corallo, ``Compact block relay.bip 152,'' \emph{[Online]. Available:
  https://github.com/bitcoin/bips/blob/master/bip-0152.mediawiki}, 2016.

\bibitem{hcb}
C.~Zhao, T.~Wang, S.~Zhang, and S.~C. Liew, ``Hcb: Enabling compact block in
  ethereum network with secondary pool and transaction prediction,'' \emph{IEEE
  Transactions on Network Science and Engineering}, vol.~11, no.~1, pp.
  1077--1092, 2024.

\bibitem{xthinblock}
P.~Tschipper, ``Buip-010 xtreme thinblocks,'' \emph{[Online]. Available:
  https://github.com/BitcoinUnlimited/BUIP/ blob/master/010.md}, 2016.

\bibitem{graphene}
A.~P. Ozisik, G.~Andresen, B.~N. Levine, D.~Tapp, G.~Bissias, and S.~Katkuri,
  ``Graphene: efficient interactive set reconciliation applied to blockchain
  propagation,'' in \emph{Proceedings of the ACM Special Interest Group on Data
  Communication}, 2019, pp. 303--317.

\bibitem{theoreticalBitcoin}
Y.~Shahsavari, K.~Zhang, and C.~Talhi, ``A theoretical model for block
  propagation analysis in bitcoin network,'' \emph{IEEE Transactions on
  Engineering Management}, 2020.

\bibitem{reducingfork}
B.~Liu, Y.~Qin, and X.~Chu, ``Reducing forks in the blockchain via
  probabilistic verification,'' in \emph{2019 IEEE 35th International
  Conference on Data Engineering Workshops (ICDEW)}.\hskip 1em plus 0.5em minus
  0.4em\relax IEEE, 2019, pp. 13--18.

\bibitem{multipleTxs}
S.~Sanghavi, B.~Hajek, and L.~Massouli{\'e}, ``Gossiping with multiple
  messages,'' \emph{IEEE Transactions on Information Theory}, vol.~53, no.~12,
  pp. 4640--4654, 2007.

\bibitem{gossip_ds}
D.~Shah, ``Gossip algorithms,'' \emph{Foundations and Trends® in Networking},
  vol.~3, no.~1, pp. 1--125, 2009.

\bibitem{stability}
A.~Gopalan, A.~Sankararaman, A.~Walid, and S.~Vishwanath, ``Stability and
  scalability of blockchain systems,'' \emph{Proceedings of the ACM on
  Measurement and Analysis of Computing Systems}, vol.~4, no.~2, pp. 1--35,
  2020.

\bibitem{gossip}
J.~Mundinger, R.~Weber, and G.~Weiss, ``Optimal scheduling of peer-to-peer file
  dissemination,'' \emph{Journal of Scheduling}, vol.~11, no.~2, pp. 105--120,
  2008.

\bibitem{bloxroute}
U.~Klarman, S.~Basu, A.~Kuzmanovic, and E.~G. Sirer, ``bloxroute: A scalable
  trustless blockchain distribution network whitepaper,'' \emph{IEEE Internet
  of Things Journal}, 2018.

\bibitem{37aa}
C.~Pinz{\'o}n and C.~Rocha, ``Double-spend attack models with time advantange
  for bitcoin,'' \emph{Electronic Notes in Theoretical Computer Science}, vol.
  329, pp. 79--103, 2016.

\bibitem{pow_1}
Q.~Qu, R.~Xu, Y.~Chen, E.~Blasch, and A.~Aved, ``Enable fair proof-of-work
  (pow) consensus for blockchains in iot by miner twins (mint),'' \emph{Future
  Internet}, vol.~13, no.~11, p. 291, 2021.

\bibitem{ethna}
T.~Wang, C.~Zhao, Q.~Yang, S.~Zhang, and S.~C. Liew, ``Ethna: Analyzing the
  underlying peer-to-peer network of ethereum blockchain,'' \emph{IEEE
  Transactions on Network Science and Engineering}, vol.~8, no.~3, pp.
  2131--2146, 2021.

\bibitem{flashbots}
``Flashbots,'' \url{https://github.com/flashbots/pm}, 2021, accessed January,
  2024.

\bibitem{docker}
C.~Anderson, ``Docker [software engineering],'' \emph{Ieee Software}, vol.~32,
  no.~3, pp. 102--c3, 2015.

\bibitem{GoEthereum}
V.~Buterin, ``Go ethereum,'' \emph{[Online]. Available:
  https://github.com/ethereum/go-ethereum/}, 2014.

\bibitem{decentralization}
A.~E. Gencer, S.~Basu, I.~Eyal, R.~Van~Renesse, and E.~G. Sirer,
  ``Decentralization in bitcoin and ethereum networks,'' in \emph{International
  Conference on Financial Cryptography and Data Security}.\hskip 1em plus 0.5em
  minus 0.4em\relax Springer, 2018, pp. 439--457.

\bibitem{ethernodes}
``Ethernodes,'' \url{https://www.ethernodes.org/}, 2019, accessed January,
  2022.

\bibitem{gossipsub}
D.~Vyzovitis, Y.~Napora, D.~McCormick, D.~Dias, and Y.~Psaras, ``Gossipsub:
  Attack-resilient message propagation in the filecoin and eth2. 0 networks,''
  \emph{arXiv preprint arXiv:2007.02754}, 2020.

\bibitem{14Sec}
J.~Xie, F.~R. Yu, T.~Huang, R.~Xie, J.~Liu, and Y.~Liu, ``A survey on the
  scalability of blockchain systems,'' \emph{IEEE Network}, vol.~33, no.~5, pp.
  166--173, 2019.

\bibitem{15aaa}
Q.~Zhou, H.~Huang, Z.~Zheng, and J.~Bian, ``Solutions to scalability of
  blockchain: A survey,'' \emph{IEEE Access}, vol.~8, pp. 16\,440--16\,455,
  2020.

\bibitem{iota}
W.~F. Silvano and R.~Marcelino, ``Iota tangle: A cryptocurrency to communicate
  internet-of-things data,'' \emph{Future Generation Computer Systems}, vol.
  112, pp. 307--319, 2020.

\bibitem{byteball}
A.~Churyumov, ``Byteball: A decentralized system for storage and transfer of
  value,'' \emph{URL https://byteball. org/Byteball. pdf}, 2016.

\bibitem{ELASTICO}
L.~Luu, V.~Narayanan, C.~Zheng, K.~Baweja, S.~Gilbert, and P.~Saxena, ``A
  secure sharding protocol for open blockchains,'' in \emph{Proceedings of the
  2016 ACM SIGSAC conference on computer and communications security}, 2016,
  pp. 17--30.

\bibitem{dang2019towards}
H.~Dang, T.~T.~A. Dinh, D.~Loghin, E.-C. Chang, Q.~Lin, and B.~C. Ooi,
  ``Towards scaling blockchain systems via sharding,'' in \emph{Proceedings of
  the 2019 international conference on management of data}, 2019, pp. 123--140.

\bibitem{rapidchain}
M.~Zamani, M.~Movahedi, and M.~Raykova, ``Rapidchain: Scaling blockchain via
  full sharding,'' in \emph{Proceedings of the 2018 ACM SIGSAC conference on
  computer and communications security}, 2018, pp. 931--948.

\bibitem{23aa}
``Danksharding,'' \url{https://ethereum.org/roadmap/danksharding}, 2024,
  accessed January, 2024.

\bibitem{polkadot}
G.~Wood, ``Polkadot: Vision for a heterogeneous multi-chain framework,''
  \emph{White Paper}, vol.~21, pp. 2327--4662, 2016.

\bibitem{casper}
V.~Buterin and V.~Griffith, ``Casper the friendly finality gadget,''
  \emph{arXiv preprint arXiv:1710.09437}, 2017.

\bibitem{25aa}
M.~Castro and B.~Liskov, ``Practical byzantine fault tolerance and proactive
  recovery,'' \emph{ACM Transactions on Computer Systems (TOCS)}, vol.~20,
  no.~4, pp. 398--461, 2002.

\bibitem{honeyBadgerBFT}
A.~Miller, Y.~Xia, K.~Croman, E.~Shi, and D.~Song, ``The honey badger of bft
  protocols,'' in \emph{Proceedings of the 2016 ACM SIGSAC conference on
  computer and communications security}, 2016, pp. 31--42.

\bibitem{dumbo}
B.~Guo, Z.~Lu, Q.~Tang, J.~Xu, and Z.~Zhang, ``Dumbo: Faster asynchronous bft
  protocols,'' in \emph{Proceedings of the 2020 ACM SIGSAC Conference on
  Computer and Communications Security}, 2020, pp. 803--818.

\bibitem{Hotstuff}
M.~Yin, D.~Malkhi, M.~K. Reiter, G.~G. Gueta, and I.~Abraham, ``Hotstuff: Bft
  consensus with linearity and responsiveness,'' in \emph{Proceedings of the
  2019 ACM Symposium on Principles of Distributed Computing}, 2019, pp.
  347--356.

\bibitem{17plasma}
J.~Poon and V.~Buterin, ``Plasma: Scalable autonomous smart contracts,''
  \emph{White paper}, pp. 1--47, 2017.

\bibitem{optimisticrollup}
K.~Floersch, ``Ethereum smart contracts in l2: Optimistic rollup. 2019,''
  \emph{URL: https://medium.
  com/plasma-group/ethereum-smart-contracts-in-l2-optimistic-rollup-2c1cef2ec537}.

\bibitem{optimistic_1}
M.~Moosavi, M.~Salehi, D.~Goldman, and J.~Clark, ``Fast and furious withdrawals
  from optimistic rollups,'' in \emph{5th Conference on Advances in Financial
  Technologies (AFT 2023)}.\hskip 1em plus 0.5em minus 0.4em\relax
  Schloss-Dagstuhl-Leibniz Zentrum f{\"u}r Informatik, 2023.

\bibitem{zkrollup}
V.~Buterin, ``On-chain scaling to potentially 500 tx/sec through mass tx
  validation,'' \emph{URL:
  https://ethresear.ch/t/on-chain-scalingto-potentially-500-tx-sec-through-mass-tx-validation/3477},
  2018.

\bibitem{a2}
T.~Neudecker and H.~Hartenstein, ``Network layer aspects of permissionless
  blockchains,'' \emph{IEEE Communications Surveys \& Tutorials}, vol.~21,
  no.~1, pp. 838--857, 2018.

\bibitem{txilm}
D.~Ding, X.~Jiang, J.~Wang, H.~Wang, X.~Zhang, and Y.~Sun, ``Txilm: Lossy block
  compression with salted short hashing,'' \emph{arXiv preprint
  arXiv:1906.06500}, 2019.

\bibitem{dino}
Z.~Hu and Z.~Xiao, ``Dino: A block transmission protocol with low bandwidth
  consumption and propagation latency,'' in \emph{IEEE INFOCOM 2022-IEEE
  Conference on Computer Communications}.\hskip 1em plus 0.5em minus
  0.4em\relax IEEE, 2022, pp. 1319--1328.

\bibitem{a3}
N.~Chawla, H.~W. Behrens, D.~Tapp, D.~Boscovic, and K.~S. Candan, ``Velocity:
  Scalability improvements in block propagation through rateless erasure
  coding,'' in \emph{2019 IEEE International Conference on Blockchain and
  Cryptocurrency (ICBC)}.\hskip 1em plus 0.5em minus 0.4em\relax IEEE, 2019,
  pp. 447--454.

\bibitem{code}
L.~Zhang, T.~Wang, and S.~C. Liew, ``Speeding up block propagation in bitcoin
  network: Uncoded and coded designs,'' \emph{Computer Networks}, vol. 206, p.
  108791, 2022.

\bibitem{Heco}
``Heco chain,'' \url{https://www.hecochain.com/developer.133bd45.pdf}, 2020,
  accessed January, 2024.

\bibitem{Sui}
T.~M. Team, ``The sui smart contracts platform,'' \emph{[Online]. Available:
  https://docs.sui.io/paper/sui.pdf}, 2022.

\bibitem{kadcast}
E.~Rohrer and F.~Tschorsch, ``Kadcast: A structured approach to broadcast in
  blockchain networks,'' in \emph{Proceedings of the 1st ACM Conference on
  Advances in Financial Technologies}, 2019, pp. 199--213.

\bibitem{urocissa}
Y.~Zhu, C.~Hua, D.~Zhong, and W.~Xu, ``Design of low-latency overlay protocol
  for blockchain delivery networks,'' in \emph{2022 IEEE Wireless
  Communications and Networking Conference (WCNC)}.\hskip 1em plus 0.5em minus
  0.4em\relax IEEE, 2022, pp. 1182--1187.

\bibitem{fring}
H.~Qiu, T.~Ji, S.~Zhao, X.~Chen, J.~Qi, H.~Cui, and S.~Wang, ``A
  geography-based p2p overlay network for fast and robust blockchain systems,''
  \emph{IEEE Transactions on Services Computing}, 2022.

\bibitem{Fiber}
``Fibre: Fast internet bitcoin relay engine,''
  \url{https://github.com/bitcoinfibre/bitcoinfibre}, 2017, accessed September,
  2022.

\bibitem{Falcon}
``Falcon - a fast bitcoin backbone,'' \url{https://www.falcon-net.org}, 2020,
  accessed January, 2024.

\bibitem{UTxO}
J.~Zahnentferner, ``Chimeric ledgers: Translating and unifying utxo-based and
  account-based cryptocurrencies,'' \emph{Cryptology ePrint Archive}, 2018.

\bibitem{MPTEthereum}
S.~Tikhomirov, ``Ethereum: state of knowledge and research perspectives,'' in
  \emph{International Symposium on Foundations and Practice of Security}.\hskip
  1em plus 0.5em minus 0.4em\relax Springer, 2017, pp. 206--221.

\bibitem{33ERC20}
F.~Vogelsteller and V.~Buterin, ``Token standard,'' \emph{[Online].Available:
  https://github.com/ethereum/EIPs/blob/master/EIPS/eip-20.md/}, 2015.

\end{thebibliography}

\begin{appendix}

\section{Missing transactions to cause low matched-blockbody probability in Ethereum}\label{trans}
In the Ethereum blockchain, there are three types of missing transactions to cause the low matched-blockbody probability. 
\begin{itemize}[left=0pt]
\item \textbf{Late Transaction:} Usually, a transaction is broadcast to nodes before the block that contains this transaction is mined by a miner. Nevertheless, some transactions may be received and included in a block by one miner early on when the transactions were just created. Due to network congestion, some nodes may not have received these transactions yet when the block arrives. In other words, these transactions are received later than the blocks that contain them.

\item \textbf{Local Transaction:}  According to the current transaction selecting and ordering algorithm,  miners include their locally generated transactions into a mined block with high priority. However, these local transactions may offer a low GAS price, and as a result, other nodes that receive them during their propagation may delete them and not store them in their local pools.

\item \textbf{Withheld Transaction:} For high profit or privacy,  some miners may not broadcast transactions with high GAS price but withhold them for the blocks that they mine \cite{mev_1}.  There are also other examples of transaction withholding. For example, in the recent FlashBots project \cite{flashbots},  some users privately send their transactions about DeFi contracts to miners without propagating them to other nodes. 
 \end{itemize}

\section{Account Model in Ethereum}\label{account}

\begin{figure}[t]
	\centering
        \includegraphics[width=8cm]{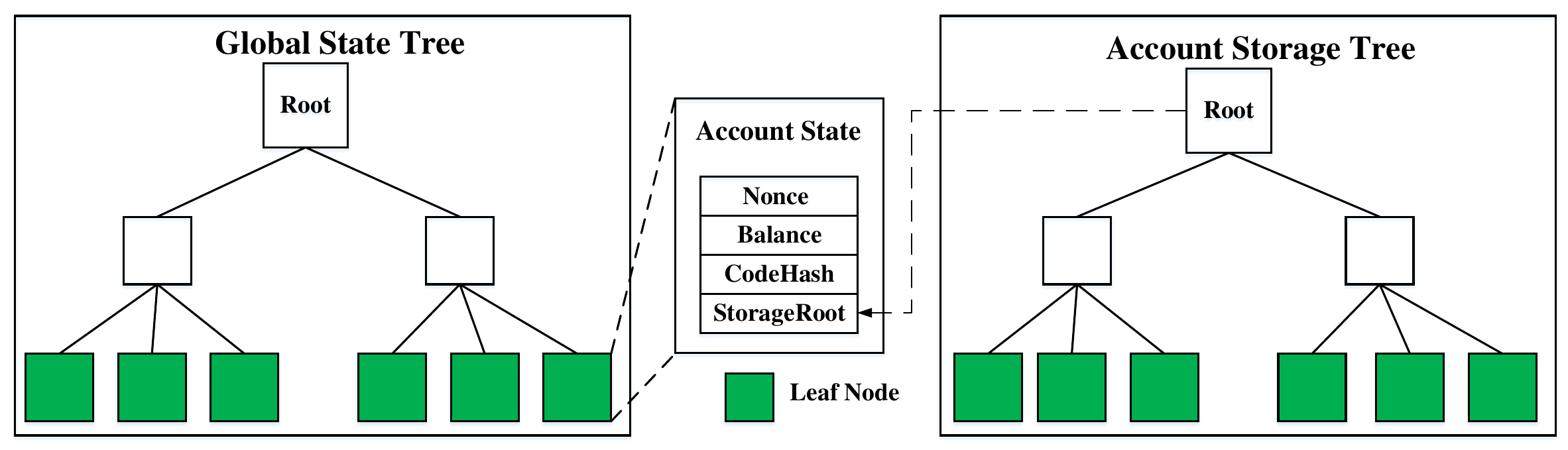}
	\caption{Illustration of account model in Ethereum.}
	\label{account_model}
\end{figure}

Blockchain is essentially a distributed database. Blockchain nodes modify the data stored in the database in a decentralized but mutually consistent manner by running a consensus algorithm. At present, there are two data storage models \cite{UTxO}, the Unspent Transaction Outputs (UTxO) model of Bitcoin and the account model of Ethereum. The UTxO model is stateless and only supports the token transfer function. By contrast, the account model is stateful and supports the execution of generic programs on blockchain called smart contracts.  Ethereum adopts the account model with smart contract functionality to extend blockchain applications.

There are two types of accounts on Ethereum: \emph{External Account} and \emph{Contract Account}.  \emph{External Accounts} are created by blockchain users to receive, hold, and send ETH, and they may invoke smart contracts to execute complex operations. \emph{Contract accounts} are used by smart contracts to store the contract code and the related data. As shown in Fig. \ref{account_model}, each account has its state and a global state tree organizes and manages the states of all accounts in the blockchain. The leaf nodes of the global state tree, indexed by the account address, record the state of their corresponding accounts. The state of the account consists of four fields: \emph{Nonce}, \emph{Balance}, \emph{CodeHash}, and \emph{StorageRoot} \cite{MPTEthereum,wood2014ethereum}.  Both the external and contract accounts have the first two fields.  \emph{Nonce} records the total number of transactions issued by the account, and each transaction also contains a nonce value corresponding to the account state.  A transaction issued by an account must be executed in ascending order of nonce\footnote{Assuming the nonce of the account state is $n$,  the nonce of a new transaction issued by this account must be set to $n+1$. If the nonce of the transaction is smaller than $n$, the transaction is called a stale transaction. If the nonce of a  transaction is larger than $n+1$, the transaction is called a future transaction. The stale transactions and future transactions cannot be propagated.}.  \emph{Balance} is the amount of ETH currently owned by the account. The remaining two fields are only related to contract accounts. \emph{CodeHash} is a 256-bit hash of the corresponding contract code.  If a transaction triggers the execution of a contract, it accesses this field to extract the contract code to complete the desired operation.  Besides \emph{CodeHash}, there is also a \emph{StorageRoot} field in the external account that points to the root state node of the account storage tree, which records all the data related to an account, e.g., the balance of ERC20 token \cite{33ERC20}.

\begin{figure}[t]
	\centering
	\includegraphics[width=8cm]{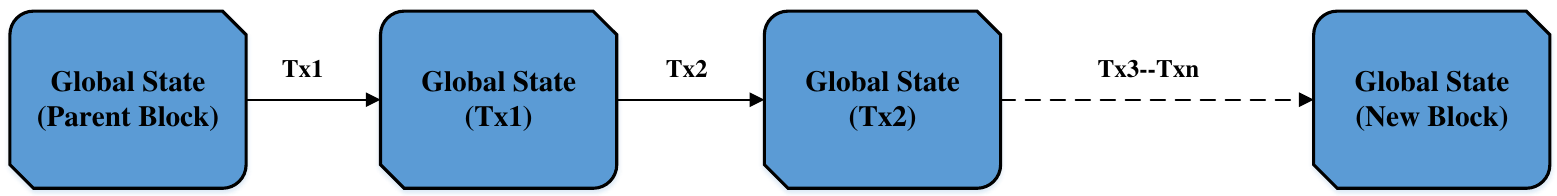}
	\caption{The process of state transition by executing transactions.}
	\label{state_process}
\end{figure}

The global state tree follows the format of the Merkle-Patricia tree \cite{wood2014ethereum}; therefore, modifying the state of the leaf node results in a different root value of the tree. The root of the global state tree is recorded in the blockheader as a snapshot of the blockchain when the miner generates the block. As shown in Fig.\ref{state_process}, upon receiving a new block, each blockchain node sequentially executes the transactions in the block using the Ethereum Virtual Machine (EVM) and modifies the global state accordingly. After executing all the transactions, the blockchain node compares the root of the new global state tree with the root in the blockheader. If the two roots match, the block is accepted and inserted into the local blockchain of the node; otherwise, the node discards the block.

\section{Transaction Selection and Ordering in Ethereum}\label{select}

In Ethereum, a blockchain node follows the Balance-and-Nonce rules\footnote{Account transfer amount and fees must smaller than the balance recorded on the global state tree, and the transaction nonce must equal to the nonce+1 recorded on the global state tree.} to issue a new transaction and propagate the transaction to other nodes on the network. When a user node receives a new transaction, it validates the transaction based on the local global state tree and the rules of the account model. If the transaction is valid, the user node stores the transaction into its local transaction pool. Before the transaction is packed by a miner node into a new block successfully, it will not be executed. Currently, there are approximately 138,000 pending transactions in the Ethereum MainNet. However, due to the limitation of the block size, each block can only include about 200 transactions, and the blocks exceeding the size limit will be rejected.  The miner node selects a batch of transactions from its local transaction pool and packs them into the block.  The transactions included in the new block and their order decide the next state of the blockchain.

\begin{figure}[t]
	\centering
	\includegraphics[width=\linewidth,height=2.5cm]{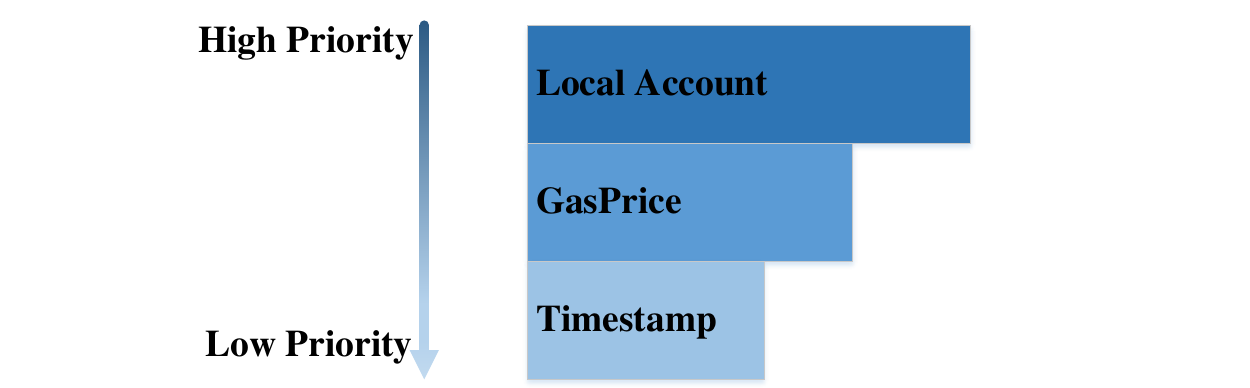}
	\caption{Priority of transactions selecting and ordering in Ethereum.}
	\label{Txs_Executing}
\end{figure}

As shown in Fig. \ref{Txs_Executing}, there are three sequential levels of priorities for transaction selection and ordering in the current Ethereum system.  For a miner node, the transactions issued by its local account enjoy the highest priority (i.e., called local transactions as discussed in Appendix \ref{trans}). In other words, to save transaction fees and accelerate its transactions, the miner node would first package its local transactions into the new block. As for the transactions issued by other nodes, the GAS price\footnote{Each transaction needs to be executed by EVM, and every computational step in EVM is priced in units of GAS. The price the user wishes to pay per unit of GAS is GAS price. The transaction fees equal the GAS price multiplied by the total GAS required to execute the transaction. That is, the larger GAS price, the larger the transaction fees.} is the key factor for transaction selection. To gain more transaction fees, the miner node usually prefers to select the transactions with the higher GAS price first. Finally, if two or more transactions have the same GAS price, the miner node selects the earliest transaction based on the timestamp.

\section{Block Generation and Validation Process in Ethereum}\label{generate}

Like Bitcoin, Ethereum also adopts the proof-of-work (PoW) consensus algorithm \cite{pow_1} to synchronize all the blockchain nodes. There are two types of Ethereum nodes: the miner nodes and the user nodes. The miner nodes compete to generate the next new block by solving a PoW puzzle. If a miner node successfully generates a new block, it then propagates the block to the other miner nodes and user nodes via the underlying P2P network. The remaining miner nodes and the user nodes perform the block validation process upon receiving the new block. Overall, the operation of the Ethereum blockchain system is a repeated process of block generation and block validation, as described below.

\begin{figure}[t]
	\centering
       \includegraphics[width=\linewidth,height=2.9cm]{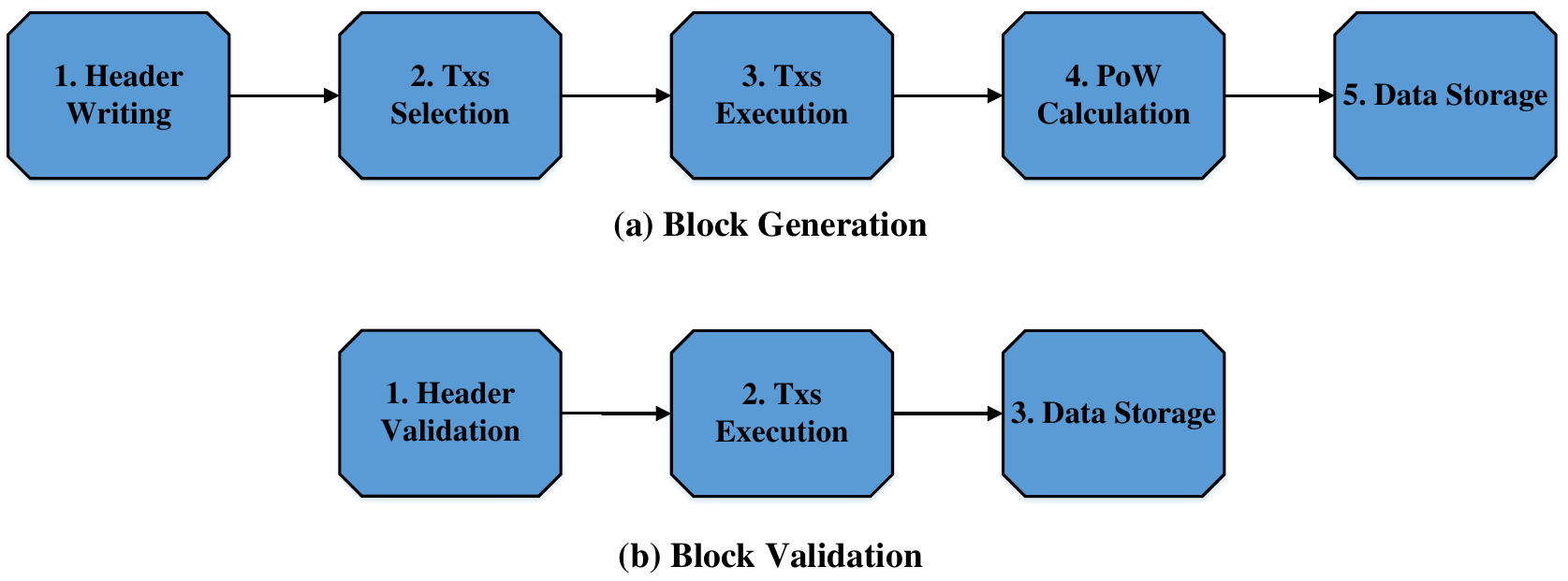}
	\caption{Process of block generation and validation.}
	\label{block_gen_validate}
\end{figure}

\noindent\textbf{\emph{Block Generation:}} Fig. \ref{block_gen_validate} illustrates the block generation process within a miner node. To generate a new block, the miner node first writes the necessary information into the block header, including the current block number, timestamp, miner address, and so on. Then, the miner selects a subset of transactions from its local transaction pool according to the prioritization discussed in the last subsection. Next, the miner node executes these transactions to update the global state tree using EVM. After that, the miner node writes the root node of the new global state tree into the block header; and then enters the mining process to calculate a valid consensus proof.  Once a valid consensus proof is successfully calculated, a legal and full block is generated.  The miner node then propagates this block to other Ethereum nodes and writes the new global state into its local database. Once the miner node receives a new block sent from another node, it suspends its block mining and enters the block validation process. 

\noindent\textbf{\emph{Block Validation:}} Fig. \ref{block_gen_validate}(b) illustrates the sequential process of block validation. The Ethereum node first validates the correctness of the blockheader. For example, it checks whether the timestamp of the new block is later than the timestamp of the parent block and whether the consensus proof is valid. If the blockheader is correct, the node sequentially executes all transactions in the block and obtains the new global state locally.  After that, the Ethereum node compares the root node of the local global state tree and the root node in the blockheader. If the two root nodes are the same, it stores the modified state and the new block in its local database. The Ethereum node discards the block if any step fails in the process of block validation.

\section{Maximum TPS of BBP}\label{maximum_tps}

As shown in Fig. \ref{basic1}(f), the relationship between the number of transactions $n_t$ in a block and the BBP block processing time $t_p$ can be fitted as 
\begin{equation}
	\setlength{\abovedisplayskip}{3pt}
	\setlength{\belowdisplayskip}{3pt}
	\label{fit_1}
	\begin{split}
		   t_p =\frac{0.398n_t + 70.725}{1000} 
	\end{split}
\end{equation}

In the scenario involving a single node without block propagation time, it holds that $t_g \geq t_p$, where $t_g$ is the block interval.   According to eq.(\ref{TPS1}) and eq.(\ref{fit_1}), the maximum TPS of BBP for one node is thereby $(1000-70.725)/0.398\approx2335$. That is, the maximum TPS in this case can be achieved by setting $n_t$ as 2,335 and $t_g$ as 1 second.

In the scenario involving multiple nodes with block propagation time, $t_g \geq t_p+t_l$, where $t_l$ is the 90\% block propagation time.  If we need to achieve this maximum TPS in the case of multiple nodes, the relationship between $n_t$ and block interval $t_g$ needs to satisfy 
\begin{equation}
	\setlength{\abovedisplayskip}{3pt}
	\setlength{\belowdisplayskip}{3pt}
	\label{fit_2}
	\begin{split}
		   \frac{n_t}{t_g}=\frac{n_t}{t_p+t_l}=2335 
	\end{split}
\end{equation}
Combining the almost constant block propagation time as shown in Fig. \ref{basic1}(c) (about 0.5 seconds) and eq.(\ref{fit_1}), eq.(\ref{fit_2}) can be rewritten as 
\begin{equation}
	\setlength{\abovedisplayskip}{3pt}
	\setlength{\belowdisplayskip}{3pt}
	\label{fit_3}
	\begin{split}
		   \frac{n_t}{t_g}=\frac{1000n_t}{0.398n_t+70.725+500}=2335 
	\end{split}
\end{equation}
From eq.(\ref{fit_3}), we can obtain $n_t \approx 18,857$ and $t_g \approx 8$ seconds. Therefore, the maximum TPS of BBP can be achieved by setting $t_g$ as 18,857 and $t_g$ as 8 seconds.

\end{appendix}
\clearpage

\end{document}